\documentclass[12pt]{article}                              
\usepackage{amsmath,amsthm,amsfonts,amscd,eucal,latexsym,amssymb}
\newlength{\dinwidth}
\newlength{\dinmargin}
\newsymbol\rest 1316         

\numberwithin{equation}{section}

\def\cA{{\cal A}}
\def\cB{{\cal B}}
\def\cC{{\cal C}}
\def\cD{{\cal D}}
\def\cE{{\cal E}}

\def\cG{{\cal G}}
\def\cH{{\cal H}}

\def\cM{{\cal M}}
\def\cN{{\cal N}}

\def\cP{{\cal P}}

\def\cR{{\cal R}}
\def\cS{{\cal S}}

\def\cU{{\cal U}}
\def\cV{{\cal V}}
\def\cW{{\cal W}}
\def\cX{{\cal X}}

\def\bC{{\mathbb C}}

\def\bN{{\mathbb N}}

\def\bR{{\mathbb R}}

\def\a{\alpha}

\def\g{\gamma}        \def\G{\Gamma}
\def\d{\delta}

\def\k{\kappa}
\def\l{\lambda}       
\def\m{\mu}
\def\n{\nu}

\def\p{\pi}
\def\r{\rho}
\def\s{\sigma}
    
\def\o{\omega}        \def\O{\Omega}

\def\fV{{\mathfrak V}}

\def\fX{{\mathfrak X}}

\def\supp{{\text{supp}}}

\newtheorem{Thm}{Theorem}[section]
\newtheorem{D+P}[Thm]{Definition and Proposition} 

\newtheorem{Prop}[Thm]{Proposition}
\newtheorem{Lemma}[Thm]{Lemma}
\theoremstyle{definition}
\newtheorem{Dfn}[Thm]{Definition}

\theoremstyle{remark}

\newsymbol\bt 1202  
\newcommand{\lcrc}{\mbox{\footnotesize $\circ$}}

\newcommand{\Cin}{C^{\infty}}
\newcommand{\Coin}{C^{\infty}_{0}}

\newcommand{\stern}{\star}

\newcommand{\fatgamma}{\boldsymbol{\gamma}}
\newcommand{\fatnabla}{\boldsymbol{\nabla}}
\newcommand{\dirop}{{\boldsymbol{\nabla}}\!\!\!\!\!/ \,}

\newcommand{\bef}{\rhd}
\newcommand{\norm}[1]{\left\lVert #1 \right\rVert}
\newcommand{\betr}[1]{\left\lvert #1 \right\rvert}
\newcommand{\scpr}[2]{\left\langle #1  ,  #2\right\rangle}
\newcommand{\scprr}[2]{\left( #1  ,  #2\right)}

\newcommand{\hide}[1]{} 
\newcommand{\ind}{\lhd}
\newcommand{\dni}{\rhd}
\newcommand{\dach}{\!\widehat{\;\;\;}}

\newcommand{\ez}{\genfrac{}{}{0pt}{1}{1}{2}} 
\newcommand{\bla}{\genfrac{}{}{0pt}{1}
{\text{terms with less than}}{\text{$m-1$ derivatives}}} 
\newcommand{\acc}[1]{\breve{#1}}

\DeclareMathOperator{\sign}{sign}
\begin{document}
\noindent
\begin{center}
{ \Large \bf Microlocal spectrum condition
  and Hadamard form\\[6pt]
 for vector-valued quantum fields in curved spacetime}
\\[30pt]
{\large \sc Hanno Sahlmann$^1$ {\rm and}  Rainer Verch$^2$}
\\[20pt]         $^1$ MPI f.\ Gravitationsphysik,\\
                     Albert-Einstein-Institut,\\
                     Am M\"uhlenberg 1,\\
                     D-14476 Golm, Germany\\[4pt]
                 e-mail: sahlmann@aei-potsdam.mpg.de\\[10pt] 
                $^2$ Institut f\"ur Theoretische Physik,\\
                 Universit\"at G\"ottingen,\\
                 Bunsenstr.\ 9,\\
                 D-37073 G\"ottingen, Germany\\[4pt]
                 e-mail: verch$@$theorie.physik.uni-goettingen.de
\end{center}
${}$\\[16pt]
{\small {\bf Abstract. }
 Some years ago, Radzikowski has found
   a characterization of Hadamard states  for
  scalar quantum fields on a four-dimensional globally hyperbolic spacetime
  in terms of a specific form of the wavefront set of their two-point
  functions (termed `wavefront set spectrum condition'),
  thereby initiating a major progress in the understanding of Hadamard
  states and the further development of quantum field theory in curved
  spacetime. In the present work, we extend this important result on
  the equivalence of the wavefront set spectrum condition with the
  Hadamard condition from scalar fields to vector
  fields (sections in a vector bundle) which are subject to a wave-equation
  and are quantized so as to fulfill the covariant canonical
  commutation relations, or which obey a Dirac equation and
  are quantized according to the covariant anti-commutation relations,
  in any globally hyperbolic spacetime having dimension three or
  higher. 

  In proving this result, a gap which is present in the published
  proof for the scalar field case will be removed. Moreover we
  determine the short-distance scaling limits of Hadamard states for
  vector-bundle valued fields, finding them to coincide with the
  corresponding flat-space, massless vacuum states.    
 }
${}$\\[10pt]
%
\section{Introduction}
%
\setcounter{equation}{0}
In quantum field theory on curved spacetime, Hadamard states have
acquired a prominent status; they are now recognized as defining the
class of physical states for quantum fields obeying linear wave
equations on any globally hyperbolic spacetime. The original
motivation for introducing Hadamard states was the observation that
they allow a definition of the expectation value of the
energy-momentum tensor with reasonable properties
\cite{Wald.Had,Ful,WaldII},
 thus Hadamard states may be viewed as a subclass of 
the states with finite energy
density. This rests basically on the fact that the two-point functions
of Hadamard states all have --- by the  very definition of Hadamard
states ---
 the same singular part which is determined
by the spacetime metric and the wave equation obeyed by the quantum
field (via the `Hadamard recursion relations') and which mimics the
singular behaviour of the vacuum state's two-point
function for linear quantum fields in flat spacetime.

A major progress in the study of Hadamard states was initiated by the
observation that, for the free scalar field, the Hadamard condition on
the two-point function of a quantum field state can be characterized
in terms of a particular, antisymmetric  form of the wavefront set of
 the two-point
function \cite{Rad1}.
 This particular form of the two-function's wavefront set is
reminiscent of the form of the support of the Fourier-transformed
two-point function of a quantum field in the vacuum state on Minkowski
spacetime and hence has been called ``wavefront set spectrum
condition'' in \cite{Rad1} and ``microlocal spectrum condition'' in
\cite{BFK}. A generalization to $n$-point functions has been suggested
in \cite{BFK}. In the present work, we will say that a state $\omega$
fulfills the microlocal spectrum condition if the wavefront set of 
its two-point function
$\omega_2$ assumes the same specific, anti-symmetric form known for
Hadamard states of a free scalar field. Expressed in formulae, this
means that the relations (5.9,5.10) in Sec.\ 5 hold.

The equivalent translation of the property of a two-point function to
be of Hadamard form into specific properties of its wavefront set made
it possible to apply the the powerful methods of microlocal analysis
(see e.g.\ the monographs
\cite{Hor1,Hor3,Tay}) in the study of Hadamard states. 
  We mention here the following results that
consequently arose:
\\[6pt] 
(a) It  has been shown that
the Hadamard form of states of the free scalar field 
is incompatible with a wide class of spacetime backgrounds which are
initially globally hyperbolic and then develop closed timelike curves
 \cite{KRW}. 
\\[6pt]
(b) ``Worldline energy inequalities'' have been established for Hadamard
states \cite{Few}. Such energy inequalities signify lower bounds for
the expectation value of the
 energy density
 integrated along
timelike curves for a suitable 
class of physical states (for instance, Hadamard states). (We refer to
\cite{Few} and the review \cite{FlWa} for further discussion and references.)
\\[6pt]
(c) A covariant definition of Wick-polynomials of the free scalar field
has been given, and generalizations of the flat space spectrum
condition to curved spacetime by a ``microlocal spectrum condition''
\cite{BFK}.
\\[6pt]
(d) A local, covariant perturbative construction of $P(\phi)_4$
theories on curved spacetime has been developed along 
the lines of
 the approach by Epstein and Glaser \cite{BF}.
\\[6pt]
 In a recent work \cite{SaV} we have shown that each ground state or
KMS-state (thermal equilibrium state) of any vector-valued quantum
field obeying a hyperbolic linear wave-equation on a stationary,
globally hyperbolic spacetime fulfills the microlocal spectrum
condition. 
The present paper may be viewed as accompanying our work
\cite{SaV}. We shall present a characterization of the Hadamard
condition for vector fields obeying a wave equation or Dirac equation
on any globally hyperbolic spacetime in terms of a specific form of
the wavefront set of the corresponding two-point functions --- in other
words, we generalize the results of \cite{Rad1} on the equivalence of
Hadamard condition and microlocal spectrum condition from the case of scalar
fields to that of vector fields and Dirac fields. Moreover, we shall
consider not only 4-dimensional spacetime, but all spacetime
dimensions $\ge 3$.

 Since two-point
functions of Hadamard states differ by a $C^{\infty}$-kernel, it is
easy to show that the results of \cite{Ver1,Ver3} generalize from the
scalar field case to the effect that all quasifree Hadamard
states of a vector-valued linear quantum field (fulfilling canonical
commutation or anti-commutation relations) induce locally unitarily
equivalent representations of the field algebra. This may provide a
starting point for generalizing the results of \cite{BF} on the
Epstein-Glaser approach to perturbative construction of interacting
quantum fields in curved spacetime from scalar fields to vector fields
which may have more direct physical relevance.

We should like to point out that, in the case of the Dirac
  field on globally hyperbolic spacetimes, results similar to ours
  have already been obtained in a couple of other works. The first of
  these is the PhD thesis by K\"ohler \cite{Koh.diss} who shows that, in
  four spacetime dimensions, the Hadamard form of the two-point
  function of quasifree states for the Dirac field can be
  characterized by the microlocal spectrum condition. This result is
  essentially the same as our Thm.\ 5.8 for the said case. In some
  more recent works, Kratzert \cite{Krat} and
Hollands \cite{Hol}  consider the Dirac
field on $n$-dimensional globally hyperbolic spacetimes. They also
present results on the equivalence of Hadamard form and microlocal
spectrum condition. Moreover, both authors
investigate also  the polarization set of the two-point functions of
Hadamard states. The polarization set is a generalization of the
wavefront set for vector-bundle distributions introduced by Dencker
\cite{Den}. In components of a local frame for a vector-bundle, a
vector-bundle distribution $u$ is locally 
represented as an element $(u_1,\ldots,u_r)$ of $\oplus^r \cD'(\bR^n)$
where $r$ is the dimension of the fibres and $n$ is the dimension of
the base-manifold (see Sec.\ 2.3). Then the elements in the
polarization set of $u$ are vectors $(x,\xi;v) \in
(T^*\bR^n\backslash\{0\}) \oplus \bC^r$ where the vectors $v$
describe, roughly speaking, which of $u$'s components has the ``most
singular'' behaviour in the microlocal sense, and $(x,\xi)$ describes
the directions of worst decay in Fourier-space of those ``most
singular'' components, like in the wavefront set. The projection of
the polarization set of $u$ onto its $(T^*\bR^n\backslash\{0\})$-part yields
the wavefront set of $u$, defined as the union of the wavefront sets
of all its components $u_1,\ldots,u_r$ which are scalar distributions.
In their works, Kratzert and Hollands determine, among other things,
 the polarization set
of the two-point functions for Hadamard states of the Dirac field and
they show that Dencker's connection, which describes the propagation of
singularities of the polarization set, coincides in this case with the
lifted spin-connection. Thereby they arrive at a characterization of
Hadamard states of the Dirac field in terms of a specific form of the
polarization sets of the corresponding two-point functions. This
characterization is somewhat more detailed than ours in terms of the
wavefront set since the polarization set contains more information
than the wavefront set. However, as is already seen from the works
\cite{Koh.diss,Krat,Hol}, the microlocal spectrum condition in terms
of the wavefront set completely characterizes the Hadamard condition
as long as the principal part of the wave operator whose wave-equation
is obeyed by the quantum field is scalar. We will exclusively consider
this case, as Hadamard forms for more general wave operators have, to
our knowledge, never been considered.

The contents of this paper are
 as follows: In Chapter 2 we summarize
the definition and basic properties of the wavefront set for scalar
and vector-bundle distributions on manifolds. This material is
included mainly to establish our notation, and to render the paper,
for the convenience of the reader, as self-contained as reasonably
possible. An auxiliary result relating the wavefront set of a
distribution to the wavefront set of its short-distance scaling limit
is also given.

Chapter 3 contains the definition of wave-operators and
Dirac-operators on vector-bundles over globally hyperbolic spacetimes
of any dimension $m \ge 3$. (Since we consider only Majorana-spinors,
there are further restrictions on $m$ in the Dirac-operator case.)
Much of the material in that chapter is patterned along the references
\cite{Dim.S,Dim.D,Keyl,Gun,V.diss}. We also quote the `propagation of
singularities theorem' for distributional solutions of wave-operators,
needed later, from \cite{DH,Den}.

In Chapter 4 we give a discussion of quantum fields obeying canonical
commutation relations (CCR) or canonical anti-commutation relations
(CAR).  We also explain how CCR- or
CAR-quantum fields are associated with wave-operators or
Dirac-operators, respectively.

In the fifth chapter we begin with the definition of Hadamard states for
vector-valued linear quantum fields obeying a wave equation or a Dirac
equation in a globally hyperbolic spacetime of dimension $ \ge 3$. Our
definition mimics the approach by Kay and Wald \cite{KW} for the scalar case,
so we are really defining ``globally Hadamard states'' whose full
definition is a bit involved.   

Then we state in Sec.\ 5.2 the result on the `propagation of Hadamard
form' in the generality needed for the present purposes and sketch the
proof, which is an entirely straightforward adaptation of the proof in
\cite{FSW} (as clarified in \cite{KW}) for the scalar field case.

In a further step, Sec.\ 5.3, we determine the short distance scaling
limits of Hadamard states which are found to coincide with the
two-point functions of the flat-space vacua for multi-component free
fields satisfying massless Klein-Gordon or Dirac equations.

Finally, we present our main result as Thm.\ 5.8 in Sec.\ 5.4,
asserting that Hadamard states of a vector-valued quantum field
satisfying a wave-equation and CCR, or a Dirac equation and CAR, can
be characterized by the specific form of the wavefront set of their
two-point functions exactly as in the scalar field case.  Prior to 
proving that result, we will point out that the original proof of
the statement for scalar fields in \cite{Rad1} contains a gap, and we
shall provide the means to complete the argument with the help of the
result on the propagation of Hadamard form.
(That gap affects also the proofs of the equivalence of
  Hadamard form and microlocal spectrum condition for Dirac fields in
  the works \cite{Koh.diss,Krat,Hol} since their authors
  rely on Radzikowski's argument.) 

Several technical issues related to Hadamard forms have been put into
the Appendix. Among them are the precise forms of Hadamard
recursion relations for wave-operators on vector-valued fields as
well as the relation of Riesz-type distributions to Hadamard forms. A
considerable part of that material has been taken from the monograph
\cite{Gun}, which we would like to advertise as a most valuable source
regarding the mathematics of Hadamard forms.
  
%
\section{On the Wavefront Set}
%

\subsection{Wavefront Sets of Scalar Distributions}
Let $n \in \bN$ and $v \in \cD'(\bR^n)$. One calls $(x,k) \in \bR^n
\times (\bR^n\backslash\{0\})$ a {\it regular directed point} for $v$
if there are $\chi \in \cD(\bR^n)$ with $\chi(x) \ne 0$, and a conical
open neighbourhood $\G$ of $k$ in $\bR^n \backslash\{0\}$ [i.e.\ $\G$
is an open neighbourhood of $k$, and $k \in \G \Leftrightarrow \m k \in
\G$ $\forall\, \m > 0$], such that
$$ \sup_{\tilde{k} \in \G}\,(1 + |\tilde{k}|)^N |\widehat{\chi
  v}(\tilde{k})| \le C_N < \infty $$
holds for all $N \in \bN$, where $\widehat{\chi v}$ denotes the
Fourier transform of the distribution $\chi\cdot v$.
\begin{Dfn}
${\rm WF}(v)$, the {\em wavefront set} of $v \in \cD'(\bR^n)$, is
defined as the complement in $\bR^n \times (\bR^n \backslash\{0\})$ of
the set of all regular directed points for $v$.
\end{Dfn}
Thus, ${\rm WF}(v)$ consists of pairs $(x,k)$ of points $x$ in
configuration space, and $k$ in Fourier space, so that the Fourier
transform of $\chi\cdot v$ isn't rapidly decaying along the direction
$k$ for large $|k|$, no matter how closely $\chi$ is concentrated
around $x$.

If $\phi : U \to U'$ is a diffeomorphism between open subsets of
$\bR^n$, and $v \in \cD'(U)$, then it holds that ${\rm WF}(\phi^*v) =
{}^tD\phi^{-1} {\rm WF}(v)$ where ${}^tD\phi^{-1}$ denotes the transpose of the
inverse tangent map (or differential) of $\phi$, with ${}^tD\phi^{-1}(x,k) =
(\phi(x),{}^tD\phi^{-1}\cdot k)$ for all $(x,k) \in {\rm WF}(v)$ and
$\phi^*v(f) = v(f \lcrc \phi)$, $f \in \cD(U')$. This
transformation behaviour of the wavefront set allows it to define the
wavefront set ${\rm WF}(v)$ of a scalar distribution
$v \in \cD'(X)$ on any
$n$-dimensional manifold $X$ [as usual, we take manifolds to be
Hausdorff, connected, 2nd countable, $C^{\infty}$ and without boundary]
by using coordinates: Let $\k : U \to \bR^n$ be a coordinate system
around a point $q \in X$. Then the dual tangent map is an isomorphism
${}^tD\k : {\rm T}_q^*X \to \bR^n$. We will use the notational
convention $(q,\xi) \in {\rm T}^*X \Leftrightarrow \xi \in {\rm
  T}_q^*X$. Then let $(q,\xi) \in {\rm T}^*X\backslash\{0\}$ and
$(x,k) := {}^tD\k^{-1}(q,\xi) = (\k(q),{}^tD\k^{-1}\cdot \xi)$, so that $(x,k)$
is in $\bR^n \times(\bR^n \backslash\{0\})$.
\begin{Dfn}
We define ${\rm WF}(v)$ by saying that $(q,\xi) \in {\rm WF}(v)$ iff
$(x,k) \in {\rm WF}(\k^*v)$ where $\k^*v$ is the chart expression of
$v$.
\end{Dfn}
Owing to the transformation properties of the wavefront set under
local diffeomorphisms one can see that this definition is independent
of the choice of the chart $\k$, and moreover, ${\rm WF}(v)$ is a
subset of ${\rm T}^*X \backslash\{0\}$, the cotangent bundle with the
zero section removed.

It is straightforward to deduce from the definition that
\begin{equation}
\label{Eq.1}
{\rm WF}(\sum_j v_j) \subset \bigcup_j{\rm WF}(v_j)
\end{equation}
for any collection of finitely many $v_1,\ldots,v_m \in \cD'(X)$, and
\begin{equation}
\label{Eq.2}
 {\rm WF}(Av) \subset {\rm WF}(v)\,, \quad v \in \cD'(X)\,,
\end{equation}
for any partial differential operator $A$ with smooth coefficients. 
(This generalizes to
pseudodifferential operators $A$.) It is also worth noting that ${\rm
  WF}(v)$ is a closed conic subset of ${\rm T}^*X \backslash\{0\}$
where conic means $(q,\xi) \in {\rm WF}(v) \Leftrightarrow (q,\m \xi)
\in {\rm WF}(v)$ $\forall\,\m > 0$. Another important property is the
following: Denote by $p_{M^*}$ the base projection of ${\rm T}^*X$,
i.e.\ $p_{M^*}: (q,\xi) \mapsto q$. Then for all $v \in \cD'(X)$ there
holds
\begin{equation}
 p_{X^*}{\rm WF}(v) = {\rm sing\,supp}\,v
\end{equation}
where sing\,supp\,$v$ is the {\em singular support} of $v$.
\begin{Dfn}
\label{D-3}
For $v \in \cD'(X)$, ${\rm sing\,supp}\,v$ is defined as the
complement of all points $q \in X$ for which there is an open
neighbourhood $U$ and a smooth $n$-form $\a_U \in \O^n(U)$ so that
$$ v(h) = \int_U h\cdot \a_U\quad {\rm for\ all}\ \ h \in \cD(U)\,.$$
\end{Dfn}
In other words, $v$ is given by an integral over a smooth $n$-form
exactly if ${\rm WF}(v)$ is empty.
\subsection{Vector Bundles and Morphisms}
Let $\fX$ be a $\Cin$ vector bundle over a base manifold $N$ (dim\,$N
=n$)
 with typical fibre  $\bC^{\,r}$ or
  $\bR^{\,r}$ and bundle
projection $\p_N$. 
The space of smooth sections of $\fX$ 
will be denoted by $\Cin(\fX)$
and $\Coin(\fX)$ denotes the
subspace of smooth sections with compact support. These spaces can be
equipped with locally convex topologies similar to those of the
test-function spaces $\cE(\bR^n)$ and $\cD(\bR^n)$, see e.g.\
\cite{Dieu3,Dieu7}. We denote by $(\Cin(\fX))'$
and $(\Coin(\fX))'$ the respective spaces of continuous
linear functionals, and by $\Coin(\fX_U)$ the space of all smooth
sections in $\fX$ with compact support in the open subset $U$ of $N$.

It will be useful to introduce the following terminology. Let $X$ be
any smooth manifold. Then 
  we say that $\r$ is a local diffeomorphism of $X$ if there are two
   open subsets $U_1 = {\rm dom}\,\r$ and $U_2 = {\rm Ran}\,\r$ of $X$ 
so that $\r : U_1 \to U_2$ is a diffeomorphism.
  If $U_1 =  U_2 = X$, then $\r$ is a diffeomorphism of $X$.
Now let $\r$ be a (local) diffeomorphism of the base manifold $N$.
Then we say that $R$ is a (local) {\it bundle map of} $\fX$
{\it covering} $\r$ if $R$ is
a smooth map from $\p_N^{-1}({\rm dom}\,\r)$ to $\p_N^{-1}({\rm
  Ran}\,\r)$ with $\p_N \lcrc R = \r$ and mapping the fibre over each
$q \in {\rm dom}\,\r$ linearly into the fibre over $\r(q)$. 
 If this map is
also one-to-one and if $R$ is also a local diffeomorphism, then $R$
will be called a (local) {\it morphism of} $\fX$ {\it covering} $\r$.

Each (local) bundle map $R$ of $\fX$ covering a (local)
diffeomorphism $\r$ of $N$ induces a (local) action on $\Coin(\fX)$,
that is, a continuous linear map $R^{\stern} : \Coin(\fX_{{\rm
    dom}\,\r}) \to \Coin(\fX_{{\rm Ran}\,\r})$ given by
\begin{equation}
 R^{\stern}f := R \lcrc f \lcrc \r^{-1}\,, \quad f \in \Coin(\fX_{{\rm
     dom}\,\r})\,.
\end{equation}
We finally note that the terminology introduced above applies
equally well to the case where $\rho$ is a local diffeomorphism
between base manifolds of \textit{different} vector bundles.   
\subsection{Wavefront Set of Vectorbundle Distributions}
Let $\fX$ again be a $C^{\infty}$ vector bundle as before. Then let $U
\subset N$ be an open subset and let $(e_1,\ldots,e_r)$ be a local
trivialization, or local frame, of $\fX$ over $U$. That means the
$e_j$, $j = 1,\ldots,r$ are sections in $\Cin(\fX_U)$ so that, for
each $q \in U$, $(e_1(q),\ldots,e_r(q))$ forms a linear basis of the
fibre $\p_N^{-1}(\{q\})$. Such a local trivialization induces a
one-to-one correspondence between $\Coin(\fX_U)$ and $\oplus^r\cD(U)$
by assigning to each $f \in \Coin(\fX_U)$ the $(f^1,\ldots,f^r) \in
\oplus^r\cD(U)$ with \footnote{summation over repeated indices is implied}
$$ f^ae_a = f\,.$$
This, in turn, induces a one-to-one correspondence between
$(\Coin(\fX_U))'$ and $\oplus^r\cD'(U)$, via mapping $u \in
(\Coin(\fX_U))'$ to the $(u_1,\ldots,u_r) \in \oplus^r\cD'(U)$ given
by
      $$ u_a(h) = u(h \cdot e_a)\,, \quad h \in \cD(U)\,. $$
With this notation, one defines for
$u \in (\Coin(\fX_U))'$ the wavefront set as
 $$ {\rm WF}(u) := \bigcup_{a = 1}^r {\rm WF}(u_a)\,,$$
i.e.\ the wavefront set of $u$ is defined as the union of the
wavefront sets of the scalar component-distributions in any local
trivialization over $U$. 
Using \eqref{Eq.1} and \eqref{Eq.2} it is straightforward to see that
this definition is independent of the choice of local trivializations.
Therefore, one is led to the following
\begin{Dfn}
Let $u \in (\Coin(\fX))'$, $(q,\xi) \in {\rm
  T}^*N\backslash\{0\}$. Then $(q,\xi)$ is defined to be in ${\rm
  WF}(u)$ if, for any neighbourhood $U$ of $q$ over which $\fX$ trivializes,
$(q,\xi)$ is in ${\rm WF}(u_U)$ where $u_U$ is the restriction of $u$
to $\Coin(\fX_U)$.
\end{Dfn}
The properties of ${\rm WF}(u)$ are similar to those in the case of
scalar distributions; obviously \eqref{Eq.1} and \eqref{Eq.2}
generalize to the vectorbundle case.
 Also, ${\rm WF}(u)$ is a closed conic
subset of ${\rm T}^*N\backslash\{0\}$,
and it holds that 
\begin{equation}
p_{N^*}{\rm WF}(u) \subset {\rm sing\,supp}\,u\,, \quad u \in
(\Coin(\fX))'\,,
\end{equation}
where $p_{N^*}: {\rm T}^*N \to N$ is the cotangent bundle projection,
and the counterpart of Def.\ \ref{D-3}  relevant here is:
\begin{Dfn}
For $u \in (\Coin(\fX))'$, ${\rm sing\,supp}\,u$ is defined as the
complement of all points in $N$ for which there are an open
neighbourhood $U$, a smooth section $\n\in \Cin(\fX^*)$ in the
dual bundle $\fX^*$ to $\fX$, and a smooth $n$-form $\a_U \in \O^n(U)$ so that
$$ u(f) = \int_U \n(f)\cdot \a_U\,, \quad f \in \Coin(\fX_U)\,.$$
\end{Dfn}
As in the scalar case, it is very useful to know the behaviour of the
wavefront set under (local) morphisms of $\fX$. The following Lemma
provides this information. The proof can be given by simply
adapting the arguments well-known for the scalar case.
\begin{Lemma} 
\label{lem3}
Let $U_1$ and $U_2$ be open subsets of $N$, and let $R :
  \fX_{U_1} \to \fX_{U_2}$ be a vector bundle map covering a
  diffeomorphism $\r: U_1 \to U_2$. Let $u \in (\Coin(\fX_{U_1}))'$.
 Then it holds that
\begin{equation}
\label{equ3-trafo}
 {\rm WF}(R^{\stern}u) \subset {}^t D\r{\rm WF}(u)
 = \{(\rho^{-1}(x),{}^t\!D\rho \cdot \xi): (x,\xi) \in {\rm WF}(u)\}\,,
\end{equation}
where ${}^t D\r$ denotes the transpose (or dual) of the tangent
map of $\r$. If $R$ is even a bundle morphism, then the inclusion
\eqref{equ3-trafo} becomes an equality.
\end{Lemma}
Note that the above Lemma applies equally well to the case of bundle
morphisms covering diffeomorphisms between base spaces of
\textit{different} vector bundles. 
\subsection{Scaling Limits}
\label{sec1}
In the present subsection we consider scaling limits of
vector-bundle distributions.

Let $\fX$ be a vector bundle with base $N$ as before. Let $q \in N$ and
let $\k : U \to O \subset \bR^n$ be a coordinate chart around $q$. We
assume that the chart range $O$ is convex and that $\k(q) = 0$. Then
we define the following semi-group $(\check{\d}_{\l})_{1>\l>0}$ of
local diffeomorphisms of $O$:
$$ \check{\d}_{\l}(y) := \l \cdot y\,, \quad y \in O,\ 1> \l >0\,.$$
This induces a semi-group $({\d}_{\l})_{1>\l>0}$ of local
diffeomorphisms of $U$ according to
$$ {\d}_{\l} = \k^{-1} \lcrc \check{\d}_{\l}
\lcrc \k\,, \quad 1 > \l > 0\,.$$
(Note that $({\d}_{\l})_{1>\l>0}$ depends on $\k$, which is not
reflected by our notation.)

Now let $(D_{\l})_{1>\l>0}$ be a family of local morphisms of $\fX$ so
that $D_{\l}$ covers $\d_{\l}$ for each $1 > \l > 0$.
\begin{Dfn}
Let $u \in (\Coin(\fX_U))'$. If the limit
$$ u^{(0)}(f) := \lim_{\l \to 0}\, u(D_{\l}^{\stern}f)$$
exists for all $f \in \Coin(\fX_U)$ and does not vanish for all $f$,
then $u^{(0)}$ will be called the {\em scaling limit distribution} with
respect to $(D_{\l})_{1> \l > 0}$ at $q$.
\end{Dfn}
Clearly, the scaling limit distribution is then a member of
$(\Coin(\fX_U))'$.
The following result will later be of interest.
\begin{Prop}
\label{P-1}
Let $u^{(0)}$ be the scaling limit distribution of a $u \in
(\Coin(\fX_U))'$ at $q$ with respect to some $(D_{\l})_{1>\l>0}$ such
that
\begin{equation}
\label{Eq.3}
\max_{1\leq a,b\leq r}\betr{D^a_b(\lambda)}\leq
c\lambda^{-\nu},\quad\quad 0<\lambda<\l_0,
\end{equation}  
holds for the components $D^a_b(\lambda)$
of $D_\lambda$ in any local trivialization of $\fX$ near $q$ with
suitable constants $c,\n > 0$.

Then
\begin{equation}
\label{Eq.4}
(q,\xi)\in{\rm WF}(u^{(0)})\Rightarrow (q,\xi) \in {\rm WF}(u). 
\end{equation}
\end{Prop}
\begin{proof}
We will establish the relation 
\begin{equation}
\label{equ28}
(q,\xi) \notin {\rm WF}(u)\Rightarrow (q,\xi)\notin{\rm WF}(u^{(0)})
\end{equation}
which is equivalent to \eqref{Eq.4}. Using the chart, we identify 
T$^*_qN$ with $\bR^n$, and we identify the components 
$(u^{(0)}_1,\ldots,u^{(0)}_r)$ of $u^{(0)}$ and $(u_1,\ldots,u_r)$ of
$u$ with respect to a local trivialization of $\fX$ near $q$ with
elements of $\cD'(O)$ where $O$ is the chart range. (Note that it is
no restriction to assume that $\fX$ trivialises on the chart domain since only the
behaviour of $u$ in any arbitrarily small neighbourhood of $q$ is
relevant here.)
With the indicated identifications provided by the chart, the required
relation \eqref{equ28} reads
\begin{equation*}
(0,\xi)\notin {\rm WF}(u_a) \text{ for all } 1\leq a\leq r
\quad\Rightarrow\quad (0,\xi) \notin {\rm WF}(u^{(0)}_a)
\text{ for all } 1\leq a\leq r 
\end{equation*}
and
\begin{equation*}
u(D^{\stern}_\lambda f)=D^a_b(\lambda)u_a(f^b\circ\delta^{-1}_\lambda)
=D^a_b(\lambda)u_a^{[\lambda]}(f^b)
\end{equation*}
in components of the local trivialization, where we have introduced 
\begin{equation*}
u_a^{[\lambda]}(f):=u_a(f\circ\delta^{-1}_\lambda),\quad\quad
f\in\cD(O).
\end{equation*} 
Now let $(0,\xi)\notin {\rm WF}(u_a)$ for all $1\leq a\leq r$. This
means that there is a function $\chi\in\Coin(O)$ with $\chi(0)=1$ and
an open conic neighbourhood $\Gamma$ of $\xi$ (in
$\bR^n\setminus\{0\}$) so that, for all $m>0$,
\begin{equation}
\label{equ30}
\sup_{k\in\Gamma}\betr{\widehat{\chi u_a}(k)}(1+\betr{k}^m)\leq C_m
\end{equation}
holds for all $1\leq a \leq r$ with suitable $C_m>0$.

Now choose some $\chi_0\in\Coin(O)$ with $\chi_0(0)=1$ and 
$\supp(\chi_0)\subset\supp(\chi)$. Then the
$\widehat{\chi_0u^{(0)}_a}(k)$ are analytic functions of $k$, hence
bounded on compact sets. This implies that it suffices to show that  
there are an open conic neighbourhood $\Gamma_0$ of $\xi$ and some
$m_0>0$ so that for all $m>m_0$ 
\begin{equation}
\label{equ29}
\sup_{k\in\Gamma_0}\betr{\widehat{\chi_0u_a^{(0)}}(k)}\betr{k}^m<C_m'
\end{equation}
holds for all $1\leq a\leq r$ with suitable $C_m'>0$, in order to
conclude that $(0,\xi)\notin{\rm WF}(u_a^{(0)})$ for all $1\leq a\leq
r$.

To prove that \eqref{equ29} holds, we observe that 
\begin{equation*}
\left((\chi u_a)^{[\lambda]}\right)\dach (k)
=\widehat{\chi u_a}(\lambda^{-1}k),\quad\quad 1>\lambda>0,\;k\in\bR^n.
\end{equation*} 
Furthermore, since \eqref{equ30} holds and since the cone $\Gamma$ is
scale-invariant, we see that 
\begin{equation*}
\sup_{1>\lambda>0,k\in\Gamma}\betr{\widehat{\chi u_a}(\lambda^{-1}k)}
\betr{\lambda^{-1}k}^m\leq C_m
\end{equation*}
for all $1\leq a\leq r$. Thus, if $m\geq\nu$, we obtain from
assumption \eqref{Eq.3}, for all $1\leq a\leq r$, 
\begin{align*}
\sup_{1>\lambda>0,k\in\Gamma}
\betr{D^b_a(\lambda)\left((\chi u_b)^{[\lambda]}\right)\dach (k)}\betr{k}^m
&\leq\sup_{1>\lambda>0,k\in\Gamma} cr^2\lambda^{-\nu}
\max_{1\leq b\leq r}\betr{\widehat{\chi u_b}(\lambda^{-1}k)}
\betr{k}^m\\
&\leq cr^2\sup_{1>\lambda>0,k\in\Gamma}\max_{1\leq b\leq r}\betr{\widehat{\chi u_b}(\lambda^{-1}k)}
\betr{\lambda^{-1}k}^m\\
&\leq cr^2C_m.
\end{align*}  
Observing that $\chi_0 u_a^{[\lambda]}=\chi_0(\chi u_a)^{[\lambda]}$
for $1>\lambda>0$ and using also that 
\begin{equation*}
(\chi_0 u_a^{(0)})\dach (k)=\lim_{\lambda\rightarrow 0}
D_a^b(\lambda)(\chi_0 u_b^{[\lambda]})\dach (k)
\end{equation*}
holds for all $k\in\bR^n$, $1\leq a\leq r$, the desired bound
\eqref{equ29} is implied by the last estimate.
\end{proof}
%
\section{Wave-operators and Dirac-operators}
%
\setcounter{equation}{0}
\subsection{Wave-operators on Vector-bundles over Curved Spacetimes}

We shall investigate the situation of general vector-valued fields
propagating over a curved spacetime. Thus, the basic object of our
considerations is a vector bundle $\fV$ with typical fibre $\bC^r$,
base manifold $M$ (dim\,$M = m$) and base projection $\p_M$. The base
manifold is to have the structure of a spacetime, so it will be
assumed that $M$ is endowed with a Lorentzian metric $g$ having
signature $(+,-,\ldots,-)$. Thus $(M,g)$ is a Lorentzian
spacetime-manifold. Within the scope of the present work, we will
impose further regularity conditions on the causal structure of
$(M,g)$. First, we assume that that $(M,g)$ is time-orientable, i.e.\
that there exists a global, timelike vectorfield on $M$. A further
assumption which we make is that $(M,g)$ be globally hyperbolic. This
means that there exists a Cauchy-surface in $(M,g)$, which by
definition is a $C^0$-hypersurface in $M$ which is intersected exactly
once by each inextendible $g$-causal curve in $M$. It can be shown
that $(M,g)$ is globally hyperbolic if and only if there exists an
$m-1$-dimensional manifold $\Sigma$ and a diffeomorphism $\Psi : \bR
\times \Sigma \to M$ so that, for each $t \in \bR$, $\Sigma_t = 
\Psi(\{t\} \times\Sigma)$ is a Cauchy-surface in $(M,g)$. 
This means that a globally
hyperbolic spacetime can be smoothly foliated by a
$C^{\infty}$-family $\{\Sigma_t\}_{t\in\bR}$ of Cauchy-surfaces.

The causal structure of globally hyperbolic spacetimes is, in a sense,
``best behaved''. In particular, it has the consequence that if $v$ is
a non-zero lightlike vector in ${\rm T}_qM$ for any $q \in M$, then
the maximal geodesic $\g$ which it defines (i.e.\ $\g: I \subset \bR
\to M$ is a solution of the geodesic equation with $\g(0) =q$ and
$\left. \frac{d}{dt}\g(t)\right|_{t=0}= v$, and any other such curve
that has the same properties cannot properly extend $\g$) is
endpointless (inextendible), and thus there is for each Cauchy-surface
 $C$ in $(M,g)$ exactly one parameter value $t$ so that $\g(t) \in C$.

Let us also collect the notation for the causal future/past sets. If
$p \in M$, then one denotes by $J^{\pm}(p)$ the subset of all points
$q$ in $M$ which lie on any future/past directed causal curve
[continuous, piecewise smooth] starting at $p$. For $G \subset M$,
$J^{\pm}(G)$ is defined as $\bigcup\{J^{\pm}(p): p \in G\}$.
For any subset $\Sigma$ of $M$ one defines its future/past domain of
dependence, $D^{\pm}(\Sigma)$, as the set of all points $p$ in $M$
such that each past/future-inextendible causal curve starting at $p$
intersects $\Sigma$ at least once. Then $D(\Sigma)$ denotes
$D^+(\Sigma) \cup D^-(\Sigma)$.   A set $G'
\subset M$ is called {\it causally separated} from $G$ if $G' \cap
(\overline{J^+(G)} \cup \overline{J^-(G)}) = \emptyset$. Note that the
relation of causal separation is symmetric in $G$ and $G'$.
The reader is referred to \cite{HE,WaldI} for a more detailed
discussion of causal structure.

After this brief reminder concerning some basic properties of
Lorentzian spacetimes, we turn now to wave-operators.
A linear partial differential operator
$$ P : \Coin(\fV) \to \Coin(\fV) $$
will be said to have {\it metric principal part} if, upon choosing a
local trivialization of $\fV$ over $U \subset M$ in which sections $f
\in \Coin(\fV_U)$ take the component representation
$(f^1,\ldots,f^r)$, and a chart $(x^{\m})_{\m =1}^m$, one has the
following coordinate representation for $P$:\,\footnote{Greek indices
  are raised and lowered with $g^{\m}{}_{\n}(x)$, latin indices with $\d^a_b$.}
$$ (Pf)^a(x) = g^{\m\n}(x)\partial_{\m}\partial_{\n}f^a(x) +
A^{\n}{}^a_b(x)\partial_{\n}f^b(x) + B^a_b(x)f^b(x)\,.$$
Here, $\partial_{\m}$ denotes the coordinate derivative
$\frac{\partial}{\partial x^{\m}}$, and $A^{\n}{}^a_b$ and $B^a_b$ are
suitable collections of smooth, complex-valued functions. Observe that
thus the principal part of $P$ diagonalizes in all local
trivializations (it is ``scalar'').

We will further suppose that there is a morphism $\G$ of $\fV$
covering the identity map of $M$ which acts as an involution ($\G
\lcrc \G = {\rm id}_{\fV}$) and operates anti-isomorphically on the
fibres. Therefore, $\G$ acts like a complex conjugation in each fibre
space, and the $\G$-invariant part $\fV^{\circ}$ of $\fV$ is a vector
bundle with typical fibre isomorphic to $\bR^r$. If $P$ has metric
principal part and is in addition $\G$-invariant, i.e.\
$$ \G \lcrc P \lcrc \G = P\,,$$
then $P$ will be called a {\it wave operator}. [Note that we have
written here $\G$ where we should have written $\G^{\stern}$, however
this appears justified since $\G$ covers the identity, so we adopt
this convention since it results in a simpler notation.]

It is furthermore worth noting that, given any wave operator, there is
a uniquely determined covariant derivative, or linear connection,
$\nabla^{(P)}$ on $\fV$, characterized by the
property
\begin{equation}
\label{nabla}
 2\cdot\nabla_{{\rm grad}\varphi}^{(P)}f = P(\varphi f) -\varphi P(f) -
(\Box \varphi)f 
\end{equation}
for all $\varphi \in \Coin(M,\bR)$ and all $f \in
\Coin(\fV^{\circ})$ \cite[Chp.\ 6]{Gun}.  
Here, $\Box$ denotes the d'Alembertian operator
associated with $g$ on the scalar functions. Then there exists also a
bundle map $V$ of $\fV^{\circ}$ covering the identity on the base manifold $M$
such that 
$$  Pf = g^{\m \n}\nabla_{\m}^{(P)}\nabla_{\n}^{(P)}f + Vf\,, \quad f 
 \in \Coin(\fV^{\circ})\,.$$
(Here we have followed our convention to denote the induced action of the
bundle map covering ${\rm id}_M$ simply by $V$ instead of $V^{\stern}$.)
\subsection{Propagation of Singularities}
We consider a wave operator $P$ for a vector bundle $\fV$ over a
spacetime manifold $(M,g)$ (for the present subsection, we need not
require that $(M,g)$ be globally hyperbolic). Let $w \in (\Coin(\fV)
\otimes \Coin(\fV))'$. Then we call $w$ a {\em bisolution
  for the wave
  operator $P$ up to $C^{\infty}$}, or, for short, {\em bisolution mod
  $C^{\infty}$}, if there are two
smooth sections $\varphi, \psi \in \Cin(\fV^* \bt \fV^*)$, where
$\fV^*$ denotes the dual bundle of $\fV$ and $\fV^* \bt \fV^*$ is the
outer tensor product bundle (this is the bundle over $M \times M$
having fibre $\fV^*_p \otimes \fV^*_q$ at $(p,q) \in M \times M$, with
the obvious projection), so that
\begin{eqnarray*}
 w(Pf \otimes f') & = & \int \varphi_{(p,q)}(f(p) \otimes
 f'(q))\,d\m(p)\,d\m(q)\,,
\\
  w(f \otimes Pf') & = & \int \psi_{(p,q)}(f(p) \otimes
 f'(q))\,d\m(p)\,d\m(q)
\end{eqnarray*}
holds for all $f,f' \in \Coin(\fV)$. Here, $d\m$ denotes the volume
measure induced by the metric $g$. In view of the fact that the
projection of ${\rm WF}(w)$ to the base manifold yields ${\rm
  sing\,supp}\,w$, one can see that, upon defining
$w^{(P)},w_{2(P)} \in (\Coin(\fV)\otimes \Coin(\fV))'$ by
\begin{eqnarray*}
 w^{(P)}(f \otimes f') & := & w(Pf \otimes f')\,,\\
 w_{(P)}(f \otimes Pf') & := & w(f \otimes Pf')\,, \quad f,f' \in
 \Coin(\fV)\,,
\end{eqnarray*}
$w$ is a bisolution for the wave operator $P$ mod $\Cin$  exactly if 
$$ {\rm WF}(w^{(P)}) = \emptyset \quad  {\rm and} \quad {\rm
  WF}(w_{(P)}) = \emptyset \,. $$

In keeping with that notation, when $w,w' \in (\Coin(\fV \bt \fV))'$
we shall also say that {\it $w$ agrees with $w'$ mod $\Cin$}, or 
$$ w = w' \quad {\rm mod} \ \Cin\,,$$
if WF$(w - w') = \emptyset$.

Let us now define the set of ``null-covectors''
\begin{equation}
\label{n-cov}
\cN := \{(q,\xi) \in {\rm T}^*M : g^{\s\r}(q)\xi_{\s}\xi_{\r} = 0\}\,.
\end{equation}
Since $(M,g)$ possesses a time orientation, 
it is useful to introduce the following two disjoint future/past-oriented
parts of $\cN$, 
\begin{equation}
\label{equ16a}
\cN_{\pm}:=\{(q,\xi)\in\cN\;\vert\;\pm\xi\bef 0 \}\,,
\end{equation}
where $\xi \bef 0$ means that the vector $\xi{}^{\m} =
g^{\m\n}\xi_{\n}$ is future-pointing and non-zero.

On the set $\cN$ one can introduce an equivalence relation as follows:
\begin{Dfn}
One defines
$$ (q,\xi)  \sim (q',\xi') $$
iff there is an affinely parametrized lightlike geodesic $\g$ with
$\g(t) = q$, $\g(t') = q'$ and
$$  g^{\s\r}(q)\xi_{\r} = (\mbox{$\left.\frac{d}{ds}\right|_{s = t}$}
\g(s))^{\s}\,, \quad
 g^{\s\r}(q')\xi'_{\r} = (\mbox{$\left.\frac{d}{ds}\right|_{s = t'}$}
\g(s))^{\s}\,. $$
That is to say, $\xi$ and $\xi'$ are co-parallel to the lightlike
geodesic $\g$ connecting the base points $q$ and $q'$, and therefore
$\xi$ and $\xi'$ are parallel transports of each other along that
geodesic.

By ${\rm B}(q,\xi) := [(q,\xi)]_{\sim}$ we will denote the equivalence
class associated with $(q,\xi) \in \cN$. 
\end{Dfn}
With this terminology, we can formulate the propagation of
singularities theorem (PST) for wave-operators, which is a consequence
of more general results of Dencker \cite{Den} together with
\cite[Lemma 6.5.5]{DH}. See also \cite{KRW} for a more elementary account.
\begin{Prop}
\label{pro2}
Let $P$ be a wave operator on $\Coin(\fV)$, and suppose that $w \in
(\Coin(\fV) \otimes \Coin(\fV))'$ is a bisolution mod $\Cin$ for $P$. 
Then there holds
$$
 {\rm WF}(w) \subset \cN \times \cN
$$
and
 $$ (q,\xi;q',\xi') \in {\rm WF}(w)   
\ \ {\rm with}\ \ \xi \ne 0\ {\rm and}\ \xi' \ne 0 \quad
\Rightarrow \quad {\rm B}(q,\xi) \times {\rm B}(q',\xi') \subset
{\rm WF}(w)\,.$$
\end{Prop} 
\subsection{Propagators and Cauchy-Problem}
\label{sec.propagators}
 As in the previous section, we assume that $\fV$ is a
vector bundle with typical fibre $\bC^r$ and base manifold $M$.
 Again,
$M$ comes endowed with a Lorentzian metric $g$ with the property that the
spacetime $(M,g)$ is time-orientable and globally hyperbolic. A
time-orientation is assumed to have been chosen. Moreover we suppose
that there is a fibre wise complex conjugation $\G$ on $\fV$, and a
wave-operator $P$ operating on $\Coin(\fV)$ and commuting with
$\G$.

An additional structure will be introduced now:
We assume that $\fV$ is a hermitean vector bundle. That is, there is a
smooth section $h$ in $\fV^* \bt \fV^*$ so that, for each $p$ in $M$,
$h_p$ is a sesquilinear form on $\fV_p$ (this form need not be
positive definite). Clearly, $h$ induces an antilinear vector-bundle
morphism $\vartheta : \fV \to \fV^*$ covering the identity via
\begin{equation} 
\label{equ6-0}
h({\rm v},{\rm w}) := (\vartheta {\rm v})({\rm w})\,, \quad {\rm
  v},{\rm w} \in \fV_q,\ q \in M\,.
\end{equation}
Then one can use $h$ to introduce a non-degenerate sesquilinear form
\begin{equation}
\label{equ6-00}
 (f,f') := \int_M h(f(q),f'(q))\, d\m (q) \,, \quad f,f' \in
\Coin(\fV)\,,
\end{equation}
on $\Coin(\fV)$. The volume form $d\m$ appearing here is that
induced by the metric $g$ on $M$.

We will furthermore  make the following assumption:
\begin{equation}
\label{equ6-a}
(Pf,f') = (f,Pf')\,, \quad f,f' \in \Coin(\fV)\,.
\end{equation}
It has been observed in \cite{Keyl,Dim.D} that under the stated
assumptions the results of \cite{Leray} imply the existence of unique
advanced and retarded fundamental solutions of $P$.
A similar statement can be deduced from \cite[Prop.\ III.4.1]{Gun}. 
 We quote this result as part (a) of the subsequent proposition from
\cite{Keyl}. Part (b) of this proposition is the statement that the
Cauchy-problem for the wave-operator is well-posed. The proof of this
statement may either be given by generalizing the classical
energy-estimate arguments as given e.g.\ in \cite{HE} for
tensor-fields to sections in vector bundles, or by using the arguments
in \cite{Dim.S} Lemma A.4 to globalize the local version of that
statement which is proven e.g.\ in \cite[Prop.\ III.5.4]{Gun}.
\begin{Prop} (a)
The wave-operator $P$ possesses unique advanced/retarded fundamental
solutions, i.e.\ there is a unique pair of (continuous) linear maps 
$$ E^{\pm}: \Coin(\fV) \to C^{\infty}(\fV) $$
such that
$$ PE^{\pm}f = E^{\pm}Pf = f \quad \ {\rm and} \ \quad {\rm
  supp}(E^{\pm}f) \subset J^{\pm}({\rm supp}\,f)\,, \quad f \in
\Coin(\fV)\,.$$
Moreover, from $\G P = P\G$ it follows that
$\G E^{\pm} = E^{\pm}\G$, and if $P$ has the
hermiticity property \eqref{equ6-a}, then it holds that 
$$ (E^{\pm}f,f') = (f,E^{\mp}f')\,, \quad f,f' \in \Coin(\fV)\,.$$
(b) Let $\Sigma$ be a Cauchy-surface in $(M,g)$, and let $n$ be 
the future-pointing
unit-normal vector field along $\Sigma$. Using Gaussian normal
coordinates, $n$ determines by geodesic transport a vector field in a
neighbourhood of $\Sigma$ (the geodesic spray of $n$) which is also
denoted by $n$. Then, given any pair $f,f' \in \Coin(\fV)$, there is
exactly one $\phi \in C^{\infty}(\fV)$ solving the Cauchy-problem for
the wave operator $P$ with data induced by $f$ and $f'$, i.e.\ $\phi$
obeys
\begin{itemize}
\item[(i)] \quad $P\phi = 0$\,,
\item[(ii)] \quad $(\phi - f)\rest \Sigma = 0$\,, \quad \ \
  $(\nabla^{(P)}_n\phi - f')\rest \Sigma = 0$\,,
\end{itemize}
where $\nabla^{(P)}$ is the connection induced by $P$.
\end{Prop}
%
\subsection{Spin Structures and Spinor Fields}
In the present section, we summarize a few basics about manifolds with
spin structure and Dirac operators, following Dimock's article
\cite{Dim.D} to large extent, however generalizing parts
therein to spacetime dimensions $\ge$ 3   while specializing
at the same time to Majorana spinors. In this context, we refer the
reader to \cite{Coq}. 

As before, we suppose that $(M,g)$ is a time-orientable, globally
hyperbolic Lorentzian spacetime of dimension $m$. Additionally, we
suppose that $(M,g)$ is ``space-orientable'', i.e.\ that each
Cauchy-surface is orientable. We suppose that time- and
space-orientations have been chosen. Then we define $F(M,g)$ as the bundle of
time- and space-oriented $g$-orthonormal frames on $M$. That is,
$F(M,g)$ consists of $m$-tuples $(v_0,v_1,\ldots,v_{m-1})_{(p)}$ of
vectors $v_{\m} \in {\rm T}_pM$, $p \in M$, such that $v_0$ is
timelike and future-pointing, $(v_1,\ldots,v_{m-1})$ is a collection
of spacelike vectors having the prescribed spatial orientation, and
$g(v_{\m},v_{\n}) = \eta_{\m \n}$ where $(\eta_{\m \n})_{\m,\n =
  0}^{m-1} ={\rm diag}(+,-,\ldots,-)$ is the $m$-dimensional
Minkowski-metric in a Lorentz frame. The base projection $\p_F:F(M,g)
\to M$ is given by $(v_0,\ldots,v_{m-1})_{(p)} \mapsto p$. $F(M,g)$
has the structure of a principal fibre bundle with structure group
SO$^{\uparrow}(1,m-1)$, where the arrow signifies that the
transformations preserve the time-orientation.

The universal covering group of SO$^{\uparrow}(1,m-1)$ is ${\rm
  Spin}^{\uparrow}(1,m-1)$. Let us denote by
$$ {\rm Spin}^{\uparrow}(1,m-1) \owns {\boldsymbol {\lambda}} \mapsto
\Lambda({\boldsymbol {\lambda}}) \in
{\rm SO}^{\uparrow}(1,m-1) $$
the 2--1 covering projection. Then a {\it spin structure} for $(M,g)$
is, by definition, a principal fibre bundle $S(M,g)$ with base manifold
$M$ ($\p_S : S(M,g) \to M$ will denote the base projection) and with structure
group ${\rm Spin}^{\uparrow}(1,m-1)$, together with a bundle-homomorphism
$\phi: S(M,g) \to F(M,g)$ preserving the base points, $\p_F \lcrc \phi =
\p_S$, and having the property that
$$ \phi \lcrc R_{\boldsymbol {\lambda}} ({\boldsymbol{s}}) =
R_{\Lambda({\boldsymbol {\lambda}})}\lcrc \phi({\boldsymbol{s}})\,, \quad {\boldsymbol{s}}
\in S(M,g)\,.$$
Here, $R_{\,.\,}$ denotes the right action of the structure group on
the corresponding principal fibre bundles. A sufficient criterion for
existence of spin-structures is that $M$ is parallelizable; this is
for instance the case for all 4-dimensional globally hyperbolic
spacetimes. 

It is known (cf.\ \cite{Coq}) that for the cases $m = 3,4,9,10$\,mod\,8
 there are
{\it Majorana algebras} $\cM(1,m-1)$,  defined as the
real-linear subalgebras of ${\tt M}(\bC^{2^{[m/2]}})$ (the algebra of
complex  ${2^{[m/2]}} \times 2^{[m/2]}$ matrices\footnote{$[m/2]$
  denotes the integer part of $m/2$}) which are generated by elements
$\{\g_{\m}: \m = 0,\ldots,m-1\}$ obeying the relations:
\\[6pt]
${}$ \hspace*{4.5cm} \quad \quad$ \g_{\m}\g_{\n} + \g_{\n}\g_{\m} = 2
\eta_{\m\n},\quad  {\rm and}\\[6pt] $
${}$ \quad \quad $\g_0^* = \g_0\,, \quad \g_k^* = -\g_k \ \ (k =
1,\ldots,m-1)\,,  
\quad  \overline{\g_{\m}} = - \g_{\m}\ \ (\m = 0,\ldots,m-1) $, 
\\[6pt]
where $(\,.\,)^*$ means taking the hermitean conjugate matrix and
$\overline{(\,.\,)}$ denotes the matrix with complex conjugate
entries. Given a Majorana algebra $\cM(1,m-1)$, one can construct a
canonical, faithful group endomorphism $\ell : {\rm
  Spin}^{\uparrow}(1,m-1) \to
\cM(1,m-1)$ so that the group multiplication is carried to the matrix
product, and with the property that
$$ \ell({\boldsymbol{\lambda}})\cdot \g_{\m} =
\Lambda({\boldsymbol{\lambda}})_{\m}^{\n}\g_{\n}\cdot \ell({\boldsymbol{\lambda}})\,.$$
Therefore $\ell$ is at the same time a  linear representation
of ${\rm Spin}^{\uparrow}(1,m-1)$ on $\bC^{2^{[m/2]}}$. Thus, given a spin
structure and a Majorana algebra, one may form the vector bundle
$$ D_{\ell}M = S(M,g) \underset{\ell}{\ltimes} \bC^{2^{[m/2]}}\,,$$
the vector bundle associated to $S(M,g)$ and the representation
$\ell$ of its structure group ${\rm Spin}^{\uparrow}(1,m-1)$ on
$\bC^{2^{[m/2]}}$. It is
called the bundle of {\it Majorana spinors} (corresponding to the Dirac
representation $\ell$ induced by $\cM(1,m-1)$). The dual bundle to
$D_{\ell}M$ will be denoted by $D^*_{\ell}M$.
\\[10pt]
{\bf Remark. } It is just a matter of convenience that we restrict our
discussion to the case of Majorana spinors. One could work with Dirac
spinors as well; then one has to introduce appropriate `doublings' of
spinor bundle and Dirac operator. Such an approach has, in the context
of quantizing Dirac fields, been favoured elsewhere
\cite{Dim.D,Hol,Koh.diss,Krat,Stro}. By employing somewhat more elaborate
notation, one may generalize our results in Chapter 5 to the slightly more
general case of Dirac spinors. 
\subsection{Dirac-operators}
The metric-induced connection $\nabla$ on $TM$  naturally gives rise to
a connection on the frame bundle $F(M,g)$. Since the Lie-algebras of
${\rm Spin}_0^{\uparrow}(1,m-1)$ and ${\rm SO}^{\uparrow}(1,m-1)$ can
 be canonically
identified, that connection on $F(M,g)$ induces a connection on
$S(M,g)$, from which one obtains a linear connection on
$D_{\ell}M$. We denote the corresponding covariant derivative operator
by $\fatnabla: C^{\infty}(TM \otimes D_{\ell}M) \to
C^{\infty}(D_{\ell}M)$, $v \otimes f \mapsto
\fatnabla_v f$, without indicating the dependence on the
representation $\ell$.

Now one can proceed exactly as in \cite{Dim.D}
and introduce the spinor-tensor $\fatgamma$, the Dirac-operator
$\dirop$  and the Dirac adjoint ${\rm u}^+$. The spinor-tensor 
$\fatgamma \in \Coin(T^*M \otimes
D_{\ell}M \otimes D^*_{\ell}M)$ is defined by requiring that its components
$\fatgamma_{\m}{}^a{}_b$ with respect to  (appropriate) local
frames are equal to the matrix elements $(\g_{\m})^a{}_b$, and the Dirac
operator is introduced by setting in frame components,
 for a local section $f = f^a e_a \in
\Coin(D_{\ell}M)$,\footnote{Note that, at this point, the indices
  $\m,\n$ are frame-indices, while elsewhere they are coordinate-indices} 
  $$ (\dirop f)^a :=  \eta^{\m\nu} \fatgamma_{\m}{}^a{}_b
  (\fatnabla_{\n}  f)^b \,.$$  
The Dirac adjoint $D_{\ell}M \owns {\rm u} \mapsto {\rm u}^+ \in
D^*_{\ell}M$ is 
a base-point preserving anti-linear bundle morphism
defined by setting $({\rm u}^+)_a =
\overline{{\rm u}^b}\g_{0\,a b}$ 
for the dual frame components. This allows to define a hermitean
structure $h$ on $D_{\ell}M$ via
$$ h({\rm u},{\rm w}) := {\rm u}^+({\rm w})\,, \quad {\rm u},{\rm w}
\in D_{\ell}M\,,$$
and thus the Dirac adjunction plays the role of $\vartheta$ of the
last section. 
 One can, moreover, introduce a
conjugation $\G$ on $D_{\ell}M$ by setting, in any frame, $(\G{\rm u})^a =
\overline{{\rm u}^a}$ for the components.
Then one finds
\begin{equation}
\label{skewconj}
h(\G{\rm u},\G{\rm w}) = -h({\rm w},{\rm u})\,,\quad {\rm u},{\rm
  w} \in D_{\ell}M\,,
\end{equation}
showing that $\G$ is a skew-conjugation for the hermitean form $h$ and
that $h$ is, while non-degenerate, not positive, but rather a
conjugate skew-symmetric form (analogous to a symplectic form). As in
the last section, $h$ induces a hermitean (now, conjugate
skew-symmetric) form on $\Coin(D_{\ell}M)$ given by
\begin{equation}
\label{equ6-01}
 (f,f') := \int_M h(f(p),f'(p))\,d\m (p) \,,\quad f,f' \in
\Coin(D_{\ell}M)\,,
\end{equation}
where again $d\m$ is the volume form induced my the metric $g$ on $M$,
and clearly $\G$ acts now as skew-conjugation with respect to this
hermitean form on $\Coin(D_{\ell}M).$

 Now if ${\sf m} \ge 0$ is a constant
(more generally, it could be a $\G$-invariant bundle map of
$D_{\ell}M$ covering the identity), one may introduce a pair of Dirac
operators
\begin{equation}
\label{equ110}
D_{\dni} := \dirop + i{\sf m} \,, \quad D_{\ind} := \dirop - i{\sf m}
\,,
\end{equation}
both of which are first-order linear partial differential operators acting on
$\Cin(D_{\ell}M)$ having the same principal part. Moreover, they have
the properties:
$$ \G D_{\dni} = -D_{\dni}\G
\,, \quad D_{\dni}D_{\ind} = D_{\ind}D_{\dni}\,, \quad {\rm and}\ \
(D_{\dni}f,f') = -(f,D_{\dni}f')\,,\ \ f,f' \in \Coin(D_{\ell}M)\,,$$
and similar relations hold when replacing $D_{\dni}$ by $D_{\ind}$.
Another property, entailed by the relations (Clifford relations)
for the generators of $\cM(1,m-1)$, is that $P = D_{\dni}D_{\ind}$ is a
wave operator on $D_{\ell}M$ which fulfills the hermiticity condition
\eqref{equ6-a}.

The following
 proposition is a trivial generalization of similar
statements in \cite{Dim.D} for the four-dimensional case; we refer to
that reference for the proof.
\begin{Prop}
\label{pro5}
{\rm \cite{Dim.D}}
  Let $D_{\dni},D_{\ind}$ be the Dirac operators on $D_{\ell}M$ defined above.
 Define
$$ S_{\dni}^{\pm} := D_{\dni}E^{\pm} \quad {\rm and} \quad 
 S_{\dni} :=  S_{\dni}^+ - S_{\dni}^-\,,$$
where $E^{\pm}$ are the advanced/retarded fundamental solutions of the
wave-operator $P = D_{\dni}D_{\ind}$.
 Then it holds that $S_{\dni}^{\pm}$ is the unique
advanced/retarded fundamental solution of $D_{\ind}$, i.e.\ the
unique continuous operator from $\Coin(\fV)$ to $\Cin(\fV)$ so that
$$ D_{\ind}S_{\dni}^{\pm}f = S_{\dni}^{\pm}D_{\ind}f = f \quad {\rm
  and} \quad {\rm supp}(S_{\dni}^{\pm}f) \subset J^{\pm}({\rm supp}\,f)\,,
\quad f \in \Coin(\fV)\,.$$
Moreover, it follows that $\G S_{\dni}^{\pm} = - S_{\dni}^{\pm} \G$,
and 
$$ (S_{\dni}f,f') =  (f,S_{\dni}f') \ \ {\rm and}\ \ 
 (f,S_{\dni}f) \ge 0\,, \quad f,f' \in \Coin(\fV)\,.$$
\end{Prop}
%
 %
\section{Quantum fields, CAR and CCR}
\setcounter{equation}{0}
The present section serves to explain what it means that a quantum
field satisfies CAR or CCR. First, however, we have to make precise the 
idea of a vector-valued quantum field on a spacetime $(M,g)$: 

Let $\fV$ be a vector bundle over the base manifold $M$, carrying a
fibrewise conjugation $\Gamma$. A quantum field is then a collection
of objects $\{\Phi,\cD,\cH\}$, where  $\cH$ is a Hilbert-space, $\cD$
is a dense subspace of $\cH$ and $\Phi$ is an operator valued
distribution having domain $\cD$. That is to say, $\Phi(f)$ is for
each $f$ in $\Coin(\fV)$ a closable operator with domain $\cD$ and $\cD$
is left invariant under application of $\Phi(f)$. 
Moreover, for all $\psi,\psi'\in\cD$, the map
\begin{equation*}
\Coin(\fV)\ni f\mapsto\scpr{\psi}{\Phi(f)\psi'}
\end{equation*}     
is in $(\Coin(\fV))'$. We also require that 
\begin{equation*}
\Phi(\Gamma f)\subset\Phi(f)^*\quad\text{for all }f\in\Coin(\fV),
\end{equation*}
where $\Phi(f)^*$ denotes the adjoint operator of $\Phi(f)$. 

Let $w$ be a distribution in $(\Coin(\fV\bt\fV))'$. One defines the
symmetric ($w^{(+)}$) and antisymmetric ($w^{(-)}$) part of $w$ by
\begin{equation*}
w^{(\pm)}(f\otimes f')=\frac{1}{2}\left(w(f\otimes f')\pm 
w(f\otimes f')\right)
\end{equation*}
and continuous linear continuation to $\Coin(\fV\bt\fV)$.  
To say that a quantum field satisfies CAR or CCR amounts to specifying
the symmetric or antisymmetric part, respectively of the two-point
functions 
\begin{equation*}
w^{(\psi)}_2(f\otimes f')=\scpr{\psi}{\Phi(f)\Phi(f')\psi}_{\cH},
\quad f,f'\in\Coin(\fV),
\end{equation*}
independently of $\psi\in\cD$, $\norm{\psi}=1$. (``c-number
commutation relations''.) To describe this more concretely, we
introduce CAR- and CCR-structures.\\[6pt]
\underline{\it CAR case:} We assume that there is a complex Hilbert-space 
$(\cV,\scpr{\cdot}{\cdot}_{\cV})$ carrying a conjugation $C$, together
with a continuous linear map 
\begin{equation*}
 q_{\cV}:\Coin(\fV) \to \cV
\end{equation*}
having a dense range, such that $C\circ q_{\cV}=q_{\cV}\circ
\Gamma$. Relative to such a {\it CAR-structure}, we say that the quantum
field $\Phi$ satisfies the CAR if
\begin{equation*}
w^{(\psi)(+)}_2(f\otimes f')=\scpr{Cq_{\cV}(f)}{q_{\cV}(f')}_{\cV},\quad f,f'\in\Coin(\fV),
\end{equation*}
holds for all unit vectors $\psi\in\cD$.\\[6pt]
\underline{\it CCR case:} Here we assume that there is a (real-linear)
symplectic space $(\cS,\sigma(\cdot,\cdot))$ and a real-linear,
symplectic map 
\begin{equation*}
q_{\cS}:\Coin(\fV^\circ)\to \cS.
\end{equation*}
Relative to this {\it CCR-structure}, we say that the quantum field $\Phi$
satisfies the CCR if 
\begin{equation*}
w^{(\psi)(-)}_2(f\otimes f')=
\sigma\left(q_{\cS}(f),q_{\cS}(f')\right),\quad f,f'\in\Coin(\fV),
\end{equation*}
holds for all unit vectors $\psi\in\cD$.
\\[10pt]
One can, instead of using quantum fields, alternatively consider
states on the Borchers-algebra \cite{Bor} over the test-section space
$\Coin(\fV)$. Since we have presented this approach in \cite{SaV}, we
won't discuss this here. Instead, we very briefly sketch the
C$^*$-algebraic variant of CAR and CCR which shows how quantum fields
may be constructed from states on C$^*$-algebras associated with CAR-
or CCR-structures. 

We begin with the CAR case. Let a CAR-structure 
$(\cV,\scpr{\cdot}{\cdot}_{\cV},C,q_{\cV})$ be given. Then one can form the
corresponding self-dual CAR-algebra $\cB(\cV,C)$ \cite{Araki},
which is the C$^*$-algebra with unit $1$ generated by a family of
elements$\{B(v):v\in\cV\}$ with the relations
\begin{itemize}
\item[(i)] $v\mapsto B(v)$ is $\bC$-linear, 
\item[(ii)] $B(v)^*=B(Cv),\quad v\in\cV$,
\item[(iii)] $B(v)^*B(w)+B(w)B(v)^*=\scpr{v}{w}_{\cV}\cdot
  1,\quad v,w\in\cV$.
\end{itemize}
(There is a unique C$^*$-norm compatible with these relations.)
Now let $\omega$ be any state, i.e. a positive ($\omega(B^*B)\geq 0$),
normalized ($\omega(1)=1$) linear functional on $\cB(\cV,C)$. Then let 
$(\pi_\omega,\cH_\omega,\Omega_\omega)$ be the corresponding
GNS-representation\footnote{
We recall here the following fact. Let $\omega$ denote a positive,
normalized linear functional on a C$^*$-algebra $\cA$ with unit
1. Then  there exists a triple
$(\pi_\omega,\cH_\omega,\Omega_\omega)$, called GNS-representation of
$\omega$ where $\cH_\omega$ is a complex Hilbert-space, $\pi_\omega$
is a $*$-representation of $\cA$ by bounded operators on $\cH_\omega$,
and $\Omega_\omega$ is a unit vector in $\cH$ which is cyclic for
$\pi_\omega$ ($\pi_\omega(\cA)\Omega_\omega$ is dense in
$\cH_\omega$), with the property that
$\omega(A)=\scpr{\Omega_\omega}{\pi_\omega(A)\Omega_\omega}$ for all
$A\in\cA$. The triple $(\pi_\omega,\cH_\omega,\Omega_\omega)$ is
unique up to unitary equivalence.}. 
It induces a quantum field $\{\Phi,\cD,\cH\}$ as follows. Take
$\cH=\cH_\omega$, and define $\Phi(f)$ by
\begin{equation}
\label{equ101}
\Phi(f):=\pi_{\omega}\left(\,B(q_{\cV}(f))\,\right), \quad f\in\Coin(\fV).
\end{equation}
 As domain $\cD$ one may take $\cP\Omega_\omega$, where
$\cP$ is the set of all polynomials in the $\Phi(f),
f\in\Coin(\fV)$. One could as well take $\cD=\cH_\omega$ since the
$\Phi(f)$ are bounded operators as consequence of the CAR. It is then
straightforward to see that this quantum field satisfies the CAR. Note that 
$\Phi$ depends on the chosen state $\omega$, and each state $\omega$
induces via \eqref{equ101} a quantum field satisfying the CAR. 

We can associate with any state $\omega$ on $\cB(\cV,C)$ its two point
function $\omega_2$, defined by 
\begin{equation*}
\omega_2(f\otimes f'):=
\scpr{\Omega_{\omega}}{\Phi(f)\Phi(f')\Omega_\omega},
\quad\quad f,f'\in\Coin(\fV), 
\end{equation*}
with $\Phi(f)$ as in \eqref{equ101}; then $\omega_2$ is an element of 
$(\Coin(\fV\bt \fV))'$.

Next, we turn to the CCR-case. Let $(\cS,\sigma,q_{\cS})$ be a
CCR-structure. Then let $\cW(\cS,\sigma)$ be the CCR- or
Weyl-algebra associated with the symplectic space $(\cS,\sigma)$. This
is the C$^*$-algebra with unit 1 generated by a family of elements 
$\{W(\phi):\phi\in\cS\}$ with relations
\begin{itemize}
\item[(i)] $W(\phi)^*=W(-\phi)=W(\phi)^{-1}$,
\item[(ii)] $W(\phi)W(\psi)=e^{-i/2 \sigma(\phi,\psi)}W(\phi+\psi)$.
\end{itemize}
(Also in this case there is a unique C$^*$-norm compatible with these
relations.) Now let $\omega$ be a state on $\cW(\cS,\sigma)$ with
corresponding GNS-representation
$(\pi_\omega,\cH_\omega,\Omega_\omega)$. This state is called
\textit{regular} if, for each $\phi\in\cS$, the unitary group 
$\bR\ni t\mapsto\pi_\omega(W(t\phi))$ is strongly
continuous. Consequently, we have for each $\phi\in\cS$ a selfadjoint
generator $R(\phi)$ so that
$\pi_\omega(W(t\phi))=\exp(itR(\phi))$. However, in order to ensure
that there is a dense common invariant domain for all
$R(\phi),\phi\in\cS$, and moreover, to obtain a quantum field, one
needs to impose a stronger regularity condition. We say that $\omega$
is \textit{$C^\infty$-regular} if, for all $N \in \bN$, the map  
\begin{equation*}
\bR^N\times\Coin(\fV^\circ)^N\ni (\vec{t},\vec{f}\,)\mapsto
\omega(W(q_{\cS}(t_1f_1))\cdots W(q_{\cS}(t_Nf_N)))
\end{equation*}
is $C^\infty$ in $\vec{t}$ and if it is, together with all partial
 $\vec{t}$-derivatives,
continuous in $\vec{f}$. Note that this requires that 
$f,f'\mapsto\sigma(q_{\cS}(f),q_{\cS}(f'))$ is continuous. 
Given a $C^\infty$-regular state $\omega$ on $\cW(\cS,\sigma)$, we
obtain a quantum field $\{\Phi,\cD,\cH\}$ from the GNS-representation 
$(\pi_\omega,\cH_\omega,\Omega_\omega)$ via setting $\cH=\cH_\omega$, 
$\cD=\cP\Omega_\omega$ where $\cP$ is the set of polynomials in the
$R(\phi),\phi\in\cS$ and
\begin{equation}
\label{equ101.1}
\Phi(f)=R(q_{\cS}(f)),\quad f\in\Coin(\fV^\circ).
\end{equation}
(Then $\Phi(f)=1/2(\Phi(f+\Gamma f)+i\Phi(i(f-\Gamma f)))$ for all 
$f\in \Coin(\fV).$)
We remark that, as in the CAR-case, the quantum field depends on the
choice of a $C^\infty$-regular state $\omega$. There exist very many
$C^\infty$-regular states for $\cW(\cS,\sigma)$ once   
$f,f'\mapsto\sigma(q_{\cS}(f),q_{\cS}(f'))$ is continuous, in
particular every quasifree state on $\cW(\cS,\sigma)$ is
$C^\infty$-regular.

Similarly as above, we associate with any $C^\infty$-regular state
$\omega$ on $\cW(\cS,\sigma)$ its two-point function, 
\begin{equation*}
\omega_2(f\otimes f'):=
\scpr{\Omega_{\omega}}{\Phi(f)\Phi(f')\Omega_\omega},
\quad\quad f,f'\in\Coin(\fV),
\end{equation*}
where $\Phi(f)$ is defined by \eqref{equ101.1}. Again $\omega_2$
induces a distribution in $(\Coin(\fV\bt \fV))'$.
\\[10pt] 
Finally, we indicate how wave-operators and Dirac operators induce
CCR-structures and CAR-structures, respectively.
\\[10pt]
\underline{\it Wave operator/CCR case:}
 We assume that $\fV$ is a hermitean vector
bundle, with typical fibre $\bC^r$, and base manifold $M$ so that
$(M,g)$ is a globally hyperbolic spacetime of dimension $m \ge
3$. Furthermore, we suppose that $P$ is a wave operator acting on the
smooth sections in the vector bundle satisfying the hermiticity
condition \eqref{equ6-a}. Let $E := E^+ - E^-$ where $E^{\pm}$ are
the unique advanced/retarded fundamental solutions of $P$, cf.\ Prop.\
3.3.
Then define
\begin{eqnarray*}
 \lefteqn{\cS := \Coin(\fV^{\circ})/{\rm ker}\,E\,, \quad
q_{\cS}(f) := f + {\rm ker}\,E\,,}\\
& & \quad \sigma(q_{\cS}(f),q_{\cS}(f'))
:= (f,Ef')\,, \quad f,f' \in \Coin(\fV^{\circ})\,,
\end{eqnarray*}
where $(\,.\,,\,.\,)$ is the hermitean form on $\Coin(\fV^{\circ})$
introduced in \eqref{equ6-00}. We call the thus defined CCR-structure
the {\it CCR structure induced by $P$}.
\\[10pt]
\underline{\it Dirac operator/CAR case: }
 Let $(M,g)$ be a globally hyperbolic
spacetime of dimension $m = 3,4,9,10 \mod 8$ and let $D_{\ell}M$ be the associated
bundle of Majorana spinors. Moreover, let $S_{\dni} = S_{\dni}^{+} -
S_{\dni}^-$ where $S_{\dni}^{\pm}$ are the advanced/retarded
fundamental solutions of the operator $D_{\ind}$, cf.\ Prop. \ref{pro5}. Then
define the {\it CAR structure induced by $D_{\ind}$} by setting
\begin{eqnarray*}
 \lefteqn{q_{\cV}: \Coin(\fV)\to \Coin(\fV)/{\rm ker}\,S_{\dni}\,, \quad
 q_{\cV}(f) := f + {\rm ker}\,S_{\dni}\,,}\\
 & & \langle
 q_{\cV}(f),q_{\cV}(f') \rangle_{\cV} = (f,S_{\dni}f')\,, \quad
 Cq_{\cV}(f) := q_{\cV}(\G f)\,,\\
& & \cV := {\rm completion \ of } \ \ \Coin(\fV)/{\rm ker}\,S_{\dni}\
\ {\rm w.r.t.}\ \ \scpr{\, . \,}{\,.\,}. 
\end{eqnarray*}
Here $(\,.\,,\,.\,)$ is the hermitean
form on $\Coin(\fV)$ introduced in \eqref{equ6-01}.

Finally, it should be noted that the quantum fields associated with
these CAR and CCR structures satisfy the Dirac-equation and the wave-equation,
respectively. That is, if $(\cS,\sigma,q_{\cS})$ is the CCR-structure
induced by the wave-operator $P$ and the quantum field $\Phi$
is defined as in \eqref{equ101.1}, then
\begin{equation}
  \Phi(Pf) = 0\,, \quad f \in \Coin(\fV)\,,
\end{equation}
and if $(\cV,\langle\,.\,,\,.\,\rangle_{\cV},q_{\cV})$ is the CAR
structure induced by the Dirac operator $D_{\ind}$, and $\Phi$ is
defined as in \eqref{equ101}, then
\begin{equation}
\Phi(D_{\ind}f) = 0\,, \quad f \in \Coin(\fV)\,.
\end{equation} 

\section{Hadamard forms, Hadamard states}
%
\subsection{Definition of Hadamard forms and Hadamard states}
\label{sec.hadamard_states}
Our next task is to give the definition of Hadamard forms and of
Hadamard states. Our definition follows that given by Kay and Wald
\cite{KW} (for bosonic fields; the formulation for fermionic fields is
an adaptation of the approach in \cite{KW} together with the notion of
Hadamard form for Dirac fields in \cite{NO} which in similar form
appeared in \cite{Koh.diss} and \cite{V.diss}).
The definition of Hadamard forms in full detail is unfortunately
somewhat laborious. We proceed by first collecting the definitions of
various notions entering into the definition of Hadamard forms;
however, we relegate the full definition of the coefficient sections
determined by the Hadamard recursion relations for the wave operator
$P$ (taken from \cite{Gun}) to the Appendix.

We suppose that we are given a hermitean vector bundle $\fV$ over a
globally hyperbolic spacetime manifold $(M,g)$ ($m:=\text{dim }M\geq 3$),
together with a wave-operator $P$ on $\Coin(\fV)$ fulfilling the 
hermiticity condition
\eqref{equ6-a}. $\G$ denotes a fibrewise conjugation on $\fV$
commuting with $P$.
\begin{itemize}
\item[(a)] {\it Causally normal related points: } A convex normal
  neighbourhood in $M$ is an open domain $U$ in $M$ such that for each
  pair of points $p,q\in U$ there is a unique geodesic segment
  contained in $U$ which connects $p$ and $q$.
We denote by $\cX$
 the set of all those $(p,q) \in M \times M$
  which are causally related and for which $J^+(p) \cap J^-(q)$ and
  $J^-(p) \cap J^+(q)$ are contained in a convex normal neighbourhood
  in $M$.
\item[(b)] {\it Causal normal neighbourhoods: } According to
  \cite{KW}, an open neighbourhood $N$ of a Cauchy-surface $\Sigma$ in
  $(M,g)$ is called a {\it causal normal neighbourhood} of $\Sigma$ if
  $\Sigma$ is a Cauchy-surface for $N$ and if for each choice of $p,q
  \in N$ with $p \in J^+(q)$, there exists a convex normal
  neighbourhood in $M$ in which $J^-(p) \cap J^+(q)$ is contained. It
  is shown in \cite{KW} that each Cauchy-surface possesses causal
  normal neighbourhoods.
\item[(c)] {\it Squared geodesic distance, Hadamard coefficient
    sections: }
There is an open neighbourhood $\cU$ of $\cX$ on which $s(p,q)$, the
squared geodesic distance between $p$ and $q$ (signed, such 
that $s(p,q)>0$ for $p,q$ spacelike, $s(p,q)\leq 0$ for $p,q$ causally
related), is well
defined and smooth. Moreover, $\cU$ may be chosen
so that there are smooth sections $U$, $V^{(n)}$ and $T^{(n)}$, 
$n \in \bN$, in  $C^{\infty}((\fV \bt \fV^*)_{\cU})$ which are
uniquely determined by
the `Hadamard recursion relations' for the wave operator $P$, see
Appendix \ref{app.koeffs} for precise definition (taken from
\cite{Gun}). These are called the {\it Hadamard coefficient sections}
for $P$. If $\cU$ has been chosen in the described way, then we
call $\cU$ a {\it regular domain}.
\item[(d)] {\it $N$-regularizing functions: } Let $N$ be a causal
  normal neighbourhood of a Cauchy-surface. A smooth function $\chi :
  N \times N \to [0,1]$ will be called {\it $N$-regularizing} if there
  is a regular domain $\cU$, and an open neighbourhood
  $\cU_* \subset N \times N$ of the set of pairs of causally related
  points in $N$ with $\overline{\cU_*} \subset \cU$, such that $\chi
  \equiv 1$ on $\cU_*$ and $\chi \equiv 0$ outside of $\cU$. The sets
  $\cU_*,\cU$ are then called the {\it domain pair} corresponding to
  $\chi$. It can be shown that $N$-regularizing functions exist, a
  proof is given in \cite{Rad1}.
\item[(e)] {\it Time-functions: } A smooth function $t : M \to \bR$ is
  called a {\it time-function} if its gradient is a future-directed
  timelike vector field normalized to 1.
\end{itemize}
We also have to define some distributions on $M$. To this end, let  
$N$ be a causal normal neighbourhood, $t$ a time-function on
$N$, and $\varepsilon>0$. Moreover, let $\chi$ be an $N$-regularizing
function with domain pair $\cU_*,\cU$. Then 
we define the smooth function (with $m = {\rm dim}\,M$) 
\begin{equation}
\label{equ101.a}
\chi G^{(1)}_{\varepsilon}(x,y):=\beta^{(1)}\chi(x,y)\left( s(x,y) -
i2\varepsilon (t(x)-t(y)) + \varepsilon^2\right)^{-\frac{m}{2}+1}
\end{equation}
with support in $N\times N$.  In case $m$ is even, we also define
\begin{equation}
\label{equ102}
\chi G^{(2)}_{\varepsilon}(x,y):=\beta^{(2)}\chi(x,y)\ln\left(s(x,y) -
i2\varepsilon(t(x) -t(y)) + \varepsilon^2\right)
\end{equation}  
where the branch cut of the logarithm is taken along the negative real
line. The constants $\beta^{(1)},\beta^{(2)}$ in the above formulas are
given by\footnote{the $\G$ appearing here denotes the
  Gamma-function, not the conjugation on the underlying vector bundle
  introduced before} 
\begin{equation*}
\beta^{(1)} = \frac{1}{2}
\begin{cases}
(-1)^{\frac{m+1}{2}}\pi^{\frac{2-m}{2}}
\left[\Gamma\left(\frac{4-m}{2}\right)\right]^{-1}
&\text{for $m$ odd}\\
-\pi^{-\frac{m}{2}}\Gamma\left(\frac{m}{2}-1\right)
&\text{for $m$ even}
\end{cases},\quad\quad
\beta^{(2)}=(-1)^{\frac{m}{2}}2^{1-m}\pi^{-\frac{m}{2}}
\left[\Gamma\left(\frac{m}{2}\right)\right]^{-1}.
\end{equation*}
We define distributions on $\Coin(M\times M)$ by
\begin{equation}
\label{equ109}
\chi G^{(\ez)} (F)=\lim_{\varepsilon\rightarrow 0+}
\iint \chi G^{(\ez)}_{\varepsilon}(p,q)F(p,q)
\text{d}\mu(p)\text{d}\mu(q)\,, \quad F\in\Coin(M\times M).
\end{equation}
For an account of some
properties of these distributions, see Appendix \ref{app.dist}.   

Now we can formulate the notion of Hadamard form:  
\begin{Dfn}
We say 
that $w \in (\Coin(\fV \bt \fV))'$
is of {\it Hadamard form on $N$} for the wave operator $P$ if there are
\begin{itemize}
\item an $N$-regularizing function $\chi$ with corresponding
  domain pair $\cU_*,\cU$ (implying that the square of the geodesic distance
  $s$ and the Hadamard coefficient sections $U$, $T^{(n)}$, $V^{(n)}$, $n \in
  \bN$, for $P$ are well-defined and smooth on $\cU$),
\item a time-function $t$,
\item for each $n \in \bN$ an $H^{(n)}\in C^n((\fV \bt \fV^*)_{N
    \times N})$ 
\end{itemize}
such that for all $f,f'\in\Coin(\fV_N)$ in case \underline{$m$ odd}:
\begin{equation*}
w(\G f\otimes f')= 
\chi G^{(1)}\left((\vartheta f) T^{(n)}f'\right)+
\int (\vartheta f)(p)H^{(n)}(p,q){f'}(q)
\;\text{d}\mu(p)\;\text{d}\mu(q), 
\end{equation*}
and for \underline{$m$ even}:
\begin{equation*}
w(\G f\otimes f')=
\chi G^{(1)}\big((\vartheta  f)  U f'\big)
+\chi G^{(2)}\big((\vartheta  f)  V^{(n)} f'\big)
+\int (\vartheta f)(p)H^{(n)}(p,q){f'}(q)
\;\text{d}\mu(p)\;\text{d}\mu(q),
\end{equation*}
where we have used abbreviations like
\begin{equation*}
((\vartheta f) T^{(n)}f')(p,q):= 
(\vartheta f)_a(p)T^{(n)a}{}_b(p,q){f'}^b(q)   
\end{equation*}
to denote the function in $\Coin(N \times N)$ resulting from
contracting $\vartheta f \otimes f'$ pointwise with $T^{(n)}$. Here,
$\vartheta$ is the antilinear base-point preserving bundle-morphism
from $\fV$ onto $\fV^*$ induced by the hermitean form as in
\eqref{equ6-0}, and we have written $\vartheta$ also in places where
we should have written $\vartheta^{\stern}$ in order to simplify notation. 
\end{Dfn}
The notion of Hadamard form seems to depend on the choice of
the time-function $t$ and the $N$-regularizing function
$\chi$; however, that turns out not to be the case.
 The difference of a distribution $w$ which is
of Hadamard form relative
to an $N$-regularizing function $\chi$ and a distribution $w'$
of Hadamard form relative to an $N$-regularizing function $\chi'$ is
$C^\infty$ because $\chi$ and $\chi'$ are both equal to 1 in a
neighbourhood of the singular support of $w$ and $w'$. Thus $w$ is
Hadamard relative to $\chi'$ and $w'$ relative to $\chi$, as well ---
a different choice of $\chi$ may be absorbed into a different choice
of the $H^{(n)}$.   
 In Appendix \ref{app.hadamard_form} we will
use an argument similar to that in \cite{KW} for the case
of scalar fields to show that, given a causal normal neighbourhood $N$ of a
Cauchy-surface, the definition of Hadamard form is independent of the
choice of the time-function $t$ that entered into the definition.

  Moreover, a solution $w$ of the wave-equation mod $\Cin$ which is of
  Hadmard form has another remarkable property, known as ``propagation
  of the Hadamard form''. We will turn to this in the subsequent section. 

We are now ready to present our definition of Hadamard states
associated with the wave operator $P$ in the CCR case.
\begin{Dfn}
\label{def1}
Let $(\cS,\sigma, q_{\sigma})$ be the CCR-structure induced by the
wave operator $P$ and $\omega$ a $C^\infty$-regular state on the
CCR-algebra $\cW(\cS,\sigma)$. We say that $\omega$ is a \textit{Hadamard
state} if there is a causal normal neighbourhood $N$ of a
Cauchy-surface in $(M,g)$ so that $\omega_2$ (the two-point function
of $\omega$) is of Hadamard form.   
\end{Dfn}
The CAR case needs slightly different assumptions. Let $(M,g)$ be a
globally hyperbolic spacetime of dimension $m=3,4,9,10 \mod 8$, and
let $D_{\ell}M$ be the corresponding bundle of Majorana spinors, and 
$D_{\dni}$, $D_{\ind}$ as in \eqref{equ110}. 
\begin{Dfn}
\label{def2}
Let $(\cV,\scpr{\,.\,}{\,.\,}_{\cV},C,q_{\cV})$ be the CAR structure
induced by $D_{\ind}$, and $\omega$ a state on the CAR-algebra
$\cB(\cV,C)$. Then we call $\omega$ a \textit{Hadamard state} if there
is a causal normal neighbourhood of a Cauchy-surface in $(M,g)$, and a
$w\in\Coin((\fV\bt \fV)_{N\times N})'$ of Hadamard form on $N$ for the
wave operator $P=D_{\dni}D_{\ind}$ so that 
\begin{equation*}
\omega_2(f\otimes f')=w(D_{\dni}f\otimes f'),\quad\quad f,f'\in\Coin(\fV_{N})
\end{equation*}
holds for the two point function $\omega_2$ of $\omega$.
\end{Dfn}
We note several things in connection with this definition. First,
these definitions of Hadamard state seem to depend on the choice of a
causal normal neighbourhood $N$, but the next section will show that
this is not the case.

Moreover, for reasons of consistency with the CCR- and CAR-structures
one has to check the following neccesary conditions.
\begin{Lemma}
\label{lem4}
$(a)$   
In the wave operator/CCR-case: Given a causal normal neighbourhood
$N$, there is a Hadamard form $w$ on $N$ for the wave operator $P$ so
that
\begin{equation}
\label{equ35}
w^{(-)}(\Gamma f\otimes f')=i\scprr{f}{Ef'},\quad\quad f,f'\in\Coin(\fV_{N}).  
\end{equation}
$(b)$
In the Dirac operator/CAR-case: Given a causal normal neighbourhood
$N$, there is a Hadamard form $w$ on $N$ for the wave operator 
$P=D_{\dni}D_{\ind}$ so that 
\begin{equation}
\label{equ36}
w^{(+)}(\Gamma D_{\dni}f\otimes f')=-i\scprr{f}{S_{\dni}f'},
\quad\quad f,f'\in\Coin(D_{\ell}M_N).  
\end{equation}
$(c)$ Let $N$ be a causal normal neighbourhood of any Cauchy-surface,
and suppose that $w \in (\Coin((\fV \bt \fV)_{N \times N})'$ is of
Hadamard form for the wave-operator $P$ on $N$. Then $w$ is a
bisolution mod $\Cin$ for $P$ on $N$.
\end{Lemma}
Drawing on results of \cite{Fri,Gun}, it will be shown in Appendix
\ref{app.cons} that Hadamard forms possess the claimed properties.

However, we should point out that these properties of Hadamard forms,
while enough for our purposes later, are really only necessary
conditions for the existence of Hadamard {\it states}. First, a further
 condition is imposed by the positivity of a state, 
$\omega(A^*A)\geq 0$, for all $A\in \cW(\cS,\sigma)$ or all
$A\in\cB(\cV,C)$. At the level of two-point functions, this implies 
\begin{equation}
\label{equ.dom}
\omega_{2}(\G f\otimes f)\omega_2(\G f'\otimes f')\geq
\betr{\omega_2(\G f\otimes f')}^2
\end{equation}
for all test-sections $f$ and $f'$, both in the CCR and CAR case. 
Moreover, two-point functions $\o_2$ are {\it proper} bisolutions ---
not just mod $\Cin$ --- for the wave-operator in the CCR case, and for
the Dirac operator in the CAR case.

Since one may construct quasifree states on $\cW(\cS,\sigma)$ or
$\cB(\cV,C)$ from two point functions, the question whether Hadamard
states exist is equivalent to the question of whether there are
Hadamard forms $w$ which are proper bisolutions of the wave-operator
or the Dirac operator satisfying \eqref{equ35} or \eqref{equ36},
respectively,     
together with the property $w(\G f\otimes f)w(\G f'\otimes f')\geq
\betr{w(\G f\otimes f')}^2$ for all test-sections $f$ and $f'$. 
That question has been answered in the affirmative for the
scalar field case in the work \cite{FNW}. The argument of \cite{FNW}
rests on a ``spacetime deformation argument'', i.e.\ the property of
any globally hyperbolic spacetime to possess a ``deformed copy''
$(\tilde{M},\tilde{g})$ which is again globally hyperbolic and
coincides with $(M,g)$ on a causal normal neighbourhood $N$ of any given
Cauchy-surface while being ultrastatic in the past of $N$. On the
ultrastatic part of $(\tilde{M},\tilde{g})$, one may construct again
the CCR-algebra of the Klein-Gordon field together with a stationary
ground state which can be shown to be Hadamard. This state induces via
the dynamics of the field (i.e.\ owing to Prop.\ 3.3) a state of the
Klein-Gordon field on $N$ and thus on $(M,g)$, and by the propagation of
Hadamard form, this state is then found to be Hadamard.

We don't see any obstruction to generalizing that method to
tensor-fields and Dirac fields; however for more general
vector fields the problem arises if the vectorbundle $\fV$ on $M$
possesses, in a suitable sense, a ``deformed copy'' $\tilde{\fV}$ on
the deformed copy $(\tilde{M},\tilde{g})$ of $(M,g)$. We won't
consider that problem in our present work. 

What seems worth mentioning is that, as we will see in the next
section, the positivity condition \eqref{equ.dom} forces the hermitean
form $h$ on $\fV$ to be positive, in the wave operator/CCR case.

\subsection{Propagation of Hadamard form}
In this section we are going to present the propagation of Hadamard
form under the dynamics of a wave-operator on sections of a vector
bundle. It is a very straightforward generalization of an analogous
result for the case of scalar fields, which has been established in a
first version by Fulling, Sweeny and Wald \cite{FSW} and, for the
present notion of ``global'' Hadamard form, by Kay and Wald
\cite{KW}. The main reason for presenting the propagation of Hadamard
form result here is that, in contrast to a claim made in \cite{Rad1}
(regarding the same issue for the scalar field case), it will turn out
to be needed in establishing the characterization of Hadamard states
in terms of the wavefront sets of their two-point functions. Below, in
the proof of Thm.\ 5.8, we will explain why the argument in \cite{Rad1}
(and similarly a related argument in \cite{Koh}) contains a gap,
which will be closed by employing the propagation of Hadamard form.

The assumptions on $P,\fV,\G,(M,g)$ etc.\ are the same as in Sec.\
5.1.
\begin{Thm} {\rm \cite{FSW,FNW,KW}} (Propagation of Hadamard form)\\[4pt]
 Let $w \in (\Coin(\fV \bt \fV))'$ be a bisolution mod $\Cin$ for the
 wave-operator $P$. Moreover, assume that there is a Cauchy-surface
 $\Sigma$ in $(M,g)$ having a causal normal neighbourhood $N$ so that
 $w$ is of Hadamard form for the wave-operator on $N$.

 Then, if $N'$ is a causal normal neighbourhood of any other
 Cauchy-surface $\Sigma'$ in $(M,g)$, $w$ is also of Hadamard form for
 the wave-operator $P$ on $N'$.
\end{Thm}
\begin{proof}
Since there is essentially no deviation from the proof of this
statement given for the scalar case in the works
\cite{FSW,FNW,KW}, we shall be content with giving only a sketch of
the proof.
\\[6pt]
Let $\Sigma'$ be another Cauchy-surface with causal normal
neighbourhood $N'$. We assume first that $\Sigma' \subset {\rm
  int\,}J^{+}(\Sigma)$. Let $\Sigma'_{\sharp}\subset \Sigma'$ 
have compact closure
and set $K' = {\rm int}\,D(\Sigma'_{\sharp}) \cap N'$. Then choose any
open, relatively compact neighbourhood $\Sigma_{\sharp}$, in $\Sigma$,
of $J^-(\Sigma_{\sharp}') \cap \Sigma$. Denote the set ${\rm
  int}\,D(\Sigma_{\sharp}) \cap N$ by $K$, and denote the set ${\rm
  int}(J^+(\Sigma) \cap J^-(\Sigma'))$ by $M(\Sigma,\Sigma')$. Then
$M(\Sigma,\Sigma')$, endowed with the appropriate restriction of $g$
as spacetime metric, is a globally hyperbolic sub-spacetime of
$(M,g)$. Thus there is a foliation $\{\Sigma_t\}_{t\in \bR}$ of
$M(\Sigma,\Sigma')$ in Cauchy-surfaces. Each $\Sigma_t$ possesses a
causal normal neighbourhood $N_t$ in $M(\Sigma,\Sigma')$, so
$\{N_t\}_{t \in \bR}$ forms an open covering of $M(\Sigma,\Sigma')$.
Now consider two causal normal neighbourhoods $\tilde{N}$ and
$\tilde{N}'$ of $\Sigma$ and $\Sigma'$ (in $M$) such that their
closures are contained in $N$ and $N'$, respectively. Then let
$$ \cC = {\rm cl}[ ({\rm int}\,D(\Sigma_{\sharp}) \cap
M(\Sigma,\Sigma')) \backslash {\rm cl}(\tilde{N} \cup
\tilde{N}')]\,.$$
We write
$\cC^{\circ}$ for the open interior of $\cC$; it is also a globally
hyperbolic sub-spacetime.
Now $\cC$ is a compact subset of $M(\Sigma,\Sigma')$ and hence there
is a finite subfamily ${N_1,\ldots,N_k}$ of $\{N_t\}_{t\in\bR}$
covering $\cC$. 
 It is not very difficult to see that one may choose
such a family with the following properties: (1)
$\Sigma_{j+1} \subset {\rm int}\,J^+(\Sigma_j)$ holds for the
corresponding Cauchy-surfaces of which the $N_j$ are causal normal
neighbourhoods; (2) For all $j =
1,\ldots,k-1$ there is some $t(j)\in\bR$ with $(N_j \cap N_{j+1}) \cap
\cC^{\circ} \supset \Sigma_{t(j)} \cap \cC^{\circ}$; (3) $N_1 \cap
\cC^{\circ}$ covers $K \cap \cC^{\circ}$ and $N_k \cap \cC^{\circ}$
covers $K' \cap \cC^{\circ}$ (by enlarging $N_1$, $N_k$, $\tilde{N}$ and
$\tilde{N}'$ if necessary).
 
Now by assumption, $w$ is of Hadamard form on $N$, and thus certainly
$w$ is of Hadamard form when restricted to $K$ (that is, $w$
restricted to $\Coin((\fV \bt \fV))_{K \times K})$). By construction,
$N_1 \cap \cC^{\circ}$ covers the part $K \cap \cC^{\circ}$ of $K$. On the
other hand, $N_1 \cap \cC^{\circ}$ is a globally hyperbolic
sub-spacetime of $M(\Sigma,\Sigma')$, so there is a Hadamard form
$w_1$ on $N_1 \cap \cC^{\circ}$. Therefore, on $K \cap \cC^{\circ}$
we have $w = w_1$ 
mod $\Cin$. Now $N \backslash {\rm cl}(\tilde{N})$ contains a
Cauchy-surface $\tilde{\Sigma}$ for $M(\Sigma,\Sigma')$ (owing to the
properties of causal normal neighbourhoods). And hence, since $w_1$ is
a Hadamard form on $N_1 \cap \cC^{\circ}$ and thus a bisolution of the
wave-operator mod $\Cin$, as likewise is $w$ by assumption, this
implies that $w = w_1$ mod $\Cin$ on ${\rm int}\,D(\tilde{\Sigma} \cap
\cC^{\circ})$, as follows by a straightforward generalization of
Lemma A.2 in \cite{FNW}. But this entails that $w = w_1$ mod $\Cin$ on all of
$N_1 \cap \cC^{\circ}$, and thus $w$ is of Hadamard form $N_1 \cap
\cC^{\circ}$. From here onwards, one iterates the just given argument to
show inductively that if $w$ is of Hadamard form on $N_j \cap
\cC^{\circ}$, then it is also of Hadamard form on $N_{j+1} \cap
\cC^{\circ}$, and hence $w$ is of Hadamard form on all $N_j \cap
\cC^{\circ}$,   
$j = 1,\ldots,k$\,; cf.\ the argument of ``Cauchy-evolution in small
steps'' in \cite{FSW}. Finally, since $N_k\cap \cC^{\circ}$
 covers the part $K' \cap
\cC^{\circ}$ of $K'$, one concludes in a like manner that $w$ is also
of Hadmard form on $K'$. And since the relatively compact set
$\Sigma_{\sharp}' \subset \Sigma'$ entering in the definition of $K'$
was arbitrary, this shows that $w$ is of Hadamard form on all of $N'$.

This establishes the statement of the Theorem for the case that
$\Sigma' \subset {\rm int}\,J^+(\Sigma)$, but it is obvious that an
analogous proof establishes the statement also in the case $\Sigma'
\subset {\rm int}\,J^-(\Sigma)$. 

Now let $\Sigma'$ be an arbitrary Cauchy-surface. Then one can choose
a Cauchy-surface $\Sigma'' \subset {\rm int}(J^-(\Sigma) \cap
J^-(\Sigma'))$. One concludes first that $w$ is of Hadamard form on
any causal normal neighbourhood $N''$ of $\Sigma''$, and then that $w$
is of Hadamard form on $N'$.
\end{proof}
\subsection{Scaling limits}
\label{sec55}
Next we shall determine the short distance scaling limits of Hadamard
forms (and thereby, of Hadamard states); this also gives in
combination with Prop.\ \ref{P-1} some first information on their
wavefront sets. Some notation needs to be introduced for this purpose. 

Let $\Omega$ be a convex normal neighbourhood of a
point $p$ in $M$ such that $\fV_\Omega$ trivializes.
$\Omega$ can be covered by (inverse) normal coordinates $\xi_{p'}$ centered at
$p'$, for any $p'\in\Omega$. The precise definition of these coordinates is
given in Appendix \ref{app.nc}. 
Fixing $p \in \Omega$, we can now define dilations 
\begin{equation}
\label{equ111}
\delta_{\lambda}(q):=\xi_{p}\left(\lambda\xi^{-1}_{p}(q)\right)\,,
\quad\quad
q\in\Omega,\,\lambda\in [0,1].
\end{equation}
Let $(e_i)_{i=1\ldots r}$ be a local frame for $\fV_\Omega$. This frame
induces a local bundle morphism $D_\lambda$ covering
$\delta_{\lambda}$ via 
\begin{equation}
\label{dil}
D_\lambda(q,v^ie_i(q)) = (\delta_{\lambda}(q),v^ie_i(\delta_{\lambda}(q)))
\end{equation}
as well as a bundle morphism 
\begin{equation*}
R:\fV_{\Omega}\to  \bR^m\times \bC^r\,,\quad (q,v^i(q)e_i(q)) \mapsto
(\xi^{-1}_{p}(q),v^i(q)b_i)
\end{equation*}
where $(b_i)_{i=1\ldots r}$ is the standard basis of
$\bC^r$. Furthermore, we can express the linear map $\left.\vartheta
  \lcrc \G\right|_p : \fV_p \to \fV^*_p$ as a matrix with respect to the
basis $(\left.e_i\right|_p)_{i=1,\ldots,r}$ of $\fV_p$ and its dual
basis in $\fV^*_p$. This matrix will be denoted by $\boldsymbol{\Theta} =
(\boldsymbol{\Theta}_{ab})_{a,b =1}^r$. Then we write
$$ \big((\boldsymbol{\Theta}R^{\stern}f)R^{\stern}f'\big)(q,q') =
 \boldsymbol{\Theta}_{ab}f^b(\xi_p(q)) f'{}^a(\xi_p(q'))\,.$$  

Now let $\alpha\in\bR$. We define an action of
the dilations on test sections by
\begin{equation*}
\left(D_\lambda^{(\alpha)}f\right)(q):=\lambda^{-\alpha}
\left(D_{\lambda}^{\stern}f\right)(q)\,,\quad\quad f\in\Coin(\fV_{\Omega})\,.
\end{equation*}
We use this action to define scaling limits for distributions as
described in Section \ref{sec1}. 

The following result gives information about the scaling limit of a Hadamard
state at $(p,p) \in M \times M$.\footnote{it is customary to call this
  also simply the ``scaling limit at $p$''; this is abuse of language
  according to the definition of scaling limit in Sec.\ 2.4: Note that
  the objects $q \in N$, $\fX$, $D^*_{\l}$ of Sec.\ 2.4 correspond to
  $(p,p) \in M \times M$, $\fV \bt \fV$, $D^{(\a)}_{\l} \otimes
  D^{(\a)}_{\l}$ here.}
 For its formulation note that $G^{(1)}_{\eta}$ stands for the
distribution $G^{(1)}$ taken with respect to the flat metric (``$g =
\eta$'') on the domain of $\xi_p$ induced by the normal coordinates at
$p$. 
 The proof of this statement will be given in Appendix \ref{app.scale}. 
\begin{Prop}
\label{pro3}
Let $\alpha_1=m/2+1$, $\alpha_2=\alpha_1-1/2$ and $\omega$ be a
quasifree Hadamard state fulfilling the CCR or CAR. For the
corresponding two-point function $\omega_2$ we have
\begin{gather*}
\text{CCR case:}\quad\quad
\lim_{\lambda\rightarrow 0}\,\omega_2\left(D^{(\alpha_1)}_\lambda
f\otimes D^{(\alpha_1)}_\lambda f'\right)= 
G^{(1)}_{\eta}\big((\boldsymbol{\Theta} R^{\stern} f) R^{\stern} f'\big)
=:\omega_2^{(C)}(f\otimes f')\\
\text{CAR case:}\quad
\lim_{\lambda\rightarrow 0}\,
\omega_2\left(D_{\dni}D^{(\alpha_2)}_\lambda f
\otimes D^{(\alpha_2)}_\lambda f'\right)=
G^{(1)}_{\eta}\big((\gamma_\mu\partial^\mu \boldsymbol{\Theta}R^{\stern} f)
R^{\stern} f'\big) 
=:\omega_2^{(A)}(f\otimes f')
\end{gather*}
\end{Prop}
The statement says that the scaling limit of the two-point function
of a Hadamard state assumes the form of the two-point function for a
multicomponent field obeying the massless Klein-Gordon equation
(CCR-case) or the massless Dirac equation (CAR) on flat Minkowski
space, apart from the appearance of the invertible matrix 
$\boldsymbol{\Theta}$.

Since we have assumed that $\omega$ is a state so that 
$\omega_2(\Gamma f \otimes f)\geq 0$, $\boldsymbol{\Theta}$   
cannot be completely arbitrary. In fact, in the CCR-case, 
$\boldsymbol{\Theta}\cdot(\G \lcrc \G_0)$ must be a positive definite
matrix, and hence the sesquilinear form $h$ must be positive
definite, i.e.\ $h$ is a fibre bundle scalar product. Here, $\G \lcrc
\G_0$ means the matrix obtained in the basis $(\left.e_i\right|_p)$
from composing the conjugation $\G:\fV_p \to \fV_p$ with the
conjugation $\G_0 :
v^i\left.e_i\right|_p \mapsto \overline{v}^i\left.e_i\right|_p$. 
\begin{Lemma}
\label{lem5}
For the above defined sesquilinear forms $\omega_2^{(C)}$ and $\omega_2^{(A)}$
on $\Coin(\fV_U)$ 
there holds
\begin{itemize}
\item[(i)] $(p,\xi;p,-\xi)\in {\rm WF}(\omega_2^{(C)})$ and
\item[(ii)] $(p,\xi;p,-\xi)\in {\rm WF}(\omega_2^{(A)})$
\\[6pt] 
for all $(p,\xi)\in\cN_-$. 
\end{itemize}
\end{Lemma}
\begin{proof}
The claim (i) is easy to see: Since $\boldsymbol{\Theta}$ is an
invertible matrix, one can use Lemma \ref{lem3} (for the case of a
bundle morphism) to reduce the proof of the statement to the scalar
case, where the claimed property is well-known
(cf.\ \cite{RS2,Rad1}\footnote{Note however that in these references $(p,\xi)$
is found to lie in $\cN_+$ due to a different sign convention in the 
definition of a Hadamard form.}). 
To prove (ii), note first that again by Lemma
\ref{lem3} it is sufficient to show $(0,v;0,-v)\in{\rm
  WF}(u^{(A)}_2)$
for each past pointing, lightlike $v$ in Minkowski space, where
\footnote{we write $y^2 =
  \eta_{\m\n} y^{\m}y^{\n}$ for the squared Minkowskian distance in
  coordinates }
\begin{equation*}
u^{(A)}_2(f\otimes f') = \lim_{\epsilon\rightarrow 0+}
\int\frac{\delta_{ab}\left(\sum_{\mu=0}^{m-1} 
\boldsymbol{\gamma}_\mu\partial_{\mu}f\right)^a(y)
{f'}^b(y')}{-(y-y')^2-2i\epsilon(y^0-{y'}^0)+\epsilon^2}\;d^my\,d^my',
\quad\quad f,f'\in\oplus^r\Coin(\bR^m).
\end{equation*}
Assume the wavefront set of $u_2^{(A)}$ were empty at the base-point $(0,0)$.
Then $u_2^{(A)}$ is $C^\infty$ near $(0,0)$ and we find that 
\begin{equation*}
\lim_{\lambda \rightarrow 0} \lambda^{-\alpha}
u_2^{(A)}(f^{[\lambda]}\otimes {f'}^{[\lambda]})=0,\quad\quad f,f'
\in\oplus^r \Coin(\bR^m)  
\end{equation*}
with $\alpha \leq 2m-1$ and $f^{[\lambda]}(y):=f(\lambda^{-1}y)$.  
But $u_2^{(A)}$ is scale
invariant, i.e.
\begin{equation*}
\lambda^{3-2m}u_2^{(A)}(f^{[\lambda]}\otimes {f'}^{[\lambda]})
=u_2^{(A)}(f\otimes f')
\end{equation*}
for all $f,f' \in\oplus^r\Coin(\bR^m)$, $1>\lambda>0$, so that we are
forced to conclude that $u_2^{(A)}(f\otimes f')=0$ for all 
$f,f' \in\oplus^r\Coin(\bR^m)$. This entails
\begin{align*}  
0&=u_2^{(A)}(\gamma_\r f\otimes f')+u_2^{(A)}
(f\otimes \gamma_\r^T f')\\
&=\lim_{\epsilon\rightarrow 0+}\int\frac{\delta_{ab}
\partial_{y^\r}f^a(y)
{f'}^b(y')}{-(y-y')^2-2i\epsilon(y^0-{y'}^0)+\epsilon^2}\;d^my\,d^my'
\end{align*}
for each $\r=0,\ldots,m-1$ and all $f,f' \in\oplus^r\Coin(\bR^m)$, which
implies
\begin{equation*}
0=\lim_{\epsilon\rightarrow 0+}\int\frac{\delta_{ab}
(-\Delta f^a)(y)(-\Delta {f'}^b)(y')}
{-(y-y')^2-2i\epsilon(y^0-{y'}^0)+\epsilon^2}\;d^my\,d^my'
\end{equation*}
for all $f,f' \in\oplus^r\Coin(\bR^m)$, where $\Delta$ is the Euclidean
Laplacian in $\bR^m$. But this is clearly a contradiction since 
$(-\Delta)\otimes (-\Delta)$ is an elliptic differential operator and
thus preserves the wavefront set of the distribution 
\begin{equation*}
f \otimes f' \mapsto
\lim_{\epsilon\rightarrow 0+}\int\frac{\delta_{ab}
f^a(y){f'}^b(y')}
{-(y-y')^2-2i\epsilon(y^0-{y'}^0)+\epsilon^2}\;d^my\,d^my'  
\end{equation*}
and this is non-empty at coinciding base points as remarked
above. Therefore, there are elements $(p,\xi;p,\xi')\in{\rm
  WF}(\omega_2^{(A)})$.
However, every such element must be of the form $\xi\in
\cN_-,\;\xi'=-\xi$ since this is so for $\omega_2^{(C)}$ and $\omega_2^{(A)}$
results from $\omega_2^{(C)}$ by application of a derivative operator. So,
there is some $(p,\xi;p,-\xi),\; \xi \in \cN_-$ in ${\rm
  WF}(\omega_2^{(A)})$. Now we use that $u_2^{(A)}$ is invariant under
spatial coordinate rotations with respect to $y=0$ (i.e. rotations in
the $y^0 = 0$-hyperplane) together with Lemma \ref{lem3} to
conclude that each   
$(p,\xi;p,-\xi), \xi\in\cN_-$, is contained in ${\rm
  WF}(\omega_2^{(A)})$.
\end{proof} 
\subsection{Main Theorem}
The following theorem generalizes the results on the
  equivalence of Hadamard form and microlocal spectrum condition,
  which have first been given by Radzikowski \cite{Rad1} for the scalar field
  case, and later by K\"ohler \cite{Koh.diss}, by Kratzert \cite{Krat}
  and by Hollands \cite{Hol} for the case of Dirac fields,
 to fields that are sections in
vector bundles, and fulfill the CCR or CAR. The arguments used are in
part taken from \cite{Rad1} with some adaptations. 
 However, we won't make use of the existence of `distinguished
parametrices' for the wave operator which was established in the
scalar field case in \cite{DH}. And, as has been mentioned before,
there is a gap in the arguments of \cite{Rad1}. A similar gap affects
Cor.\ 1 in \cite{Koh}, and it also affects the statements in
\cite{Koh.diss,Krat,Hol} regarding the equivalence of Hadamard form
and microlocal spectrum condition since the authors of these works
rely on Radzikowski's main argument (as we shall also mostly do).
 We will explain in Remark (iii) below the statement of
the next theorem where this gap occurs, and will repair it in our
proof.

We also mention that the approach taken in the 
references \cite{Koh.diss,Krat,Hol} is
slightly more general to the extent that, in contrast to our approach,
it is not assumed in these references that the Dirac fields are
Majorana fields (cf.\ the Remark at the end of Sec.\ 3.4). Thus, in
these references Hadamard states are not automatically
charge-conjugation invariant in the sense that $\o_2(\G f,\G f') =
\o_2(f',f)$, as is the case here. That situation could be obtained,
however, by considering appropriate `doublings' of the field systems
considered in the mentioned references.

The assumptions are the same as in Sec.\ 5.1.
\begin{Thm}
\label{the2}
Let $\omega$ be either: 
\begin{itemize}
\item a $C^\infty$-regular state on the CCR-algebra $\cW(\cS,\sigma)$
associated with a CCR-structure induced by a wave operator $P$.  
\item a state on the CAR-algebra $\cB(\cV,C)$ associated with a
  CAR-structure induced by a Dirac operator $D_{\ind}$.
\end{itemize}
Then it holds that: 
\begin{itemize}
\item[(a)] If there is a causal normal neighbourhood $N$ of a
  Cauchy-surface in $(M,g)$ so that $\o$ is a Hadamard state 
  on $N$, then 
\begin{equation}
\label{6-WF}
{\rm WF}(\o_2) = \cR
\end{equation}
where
\begin{equation}
\label{6R-WF}
  \cR = \{(q,\xi;q',\xi') \in \cN_- \times \cN_+ : (q,\xi)
\sim (q',-\xi')\}\,.
\end{equation}
\item[(b)] Conversely, if \eqref{6-WF} holds, then $\o$ is a global Hadamard
  state. 
\end{itemize}
\end{Thm}\noindent
{\it Remark. } (i) As will be obvious from the proof, one also has the
following slightly more general statement of part (a): Let $w$ be a
bisolution mod $\Cin$ for the wave-operator, and suppose that $w$ is
of Hadamard form on $N$. Then WF$(w) = \cR$. It is not clear, however,
if the converse direction (b) holds for $w$ unless its symmetric or
antisymmetric part is suitably fixed (mod $\Cin$) as is the case for two-point
functions of quantum fields fulfilling CAR or CCR.\\[6pt]
(ii) It will also be apparent from the proof that (b) holds also under
the assumption WF$(\o_2) \subset \cR$ (and even under the
seemingly much weaker assumption WF$(\o_2) \subset \cN_- \times
\cN_+$). This proves the claim made in \cite{SaV} that a two-point
function $\o_2$ of a quantum field fulfilling CAR or CCR and WF$(\o_2)
\subset \cR$ is of Hadamard form and thus WF$(\o_2) = \cR$.\\[6pt]
 (iii) The proof of part (a) needs an argument proving that the
relation ${\rm WF}(\omega_{2\,N\times N}) \subset \cR$ implies ${\rm
  WF}(\omega_2) \subset \cR$, where $\omega_{2\,N \times N}$ denotes
the restriction of $\omega_2$ to $\Coin((\fV\bt\fV)_{N\times N})$. To
show this one invokes, as in \cite{Rad1}, the propagation of
singularities theorem which says that $(q,\xi;q',\xi') \in {\rm
  WF}(\omega_2)$ implies ${\rm B}(q,\xi) \times {\rm B}(q',\xi') \in
{\rm WF}(\omega_2)$. The argument proving the said implication
requires, however, that both bicharacteristics ${\rm B}(q,\xi)$ and
${\rm B}(q',\xi')$ really consist of inextendible lightlike geodesics,
and this is not the case if either $\xi = 0$ or $\xi' =0$ since ${\rm
  B}(q,0)$ equals $\{q\}$. Thus, when considering e.g.\ $(q,\xi;q',0)$
with $q'$ not in $N$, then ${\rm B}(q',0)$ won't meet $N$ and so one
cannot use the propagation of singularities theorem to decide if
$(q,\xi;q',0)$ is in ${\rm WF}(\omega_2)$ by knowing that ${\rm
  WF}(\omega_{2\,N\times N}) \subset \cR$. Due to having overlooked
this gap, it has been claimed explicitly in \cite{Rad1} that the
propagation of Hadamard form result were not needed in order to
conclude that ${\rm WF}(\omega_2) \subset \cR$ once it is known that
$\omega_2$ is of Hadamard form on some causal normal neighborhood $N$
of an arbitrary Cauchy surface. As far as we can see, however, the
result on the propagation of Hadamard form is needed in order to
conclude that pairs $(q,\xi;q',0)$ or $(q,0;q',\xi')$ aren't contained
in ${\rm WF}(\omega_2)$. At least it will prove sufficient to reach at
this conclusion. 
\begin{proof}
(a)  Let the element $\cG_n$ of $(\Coin((\fV\bt\fV)_{N\times N}))'$ be 
defined by 
\begin{equation*}
\cG_n(f\otimes f')=\begin{cases}
\chi G^{(1)}\big((\vartheta f) T^{(n)}f'\big) & \text{ for $m$ odd}\\
\chi G^{(1)}\big((\vartheta f) U f'\big)+ 
\chi G^{(2)}\big((\vartheta f) V^{(n)}f'\big) & \text{ for $m$ even}\,,
\end{cases}
\end{equation*}
where $\chi$ is an $N$-regularizing function. 
Using the arguments of part (i) of the proof of Thm.\ 5.1 in
\cite{Rad1} in combination with Lemma \ref{lem3}, it is 
straightforward to deduce 
${\rm WF}(\cG_n)\subset\cR\cap({\rm T}^*N\times{\rm T}^*N)$.
Thus, if $w\in(\Coin((\fV\bt\fV)_{N\times N}))'$ denotes a Hadamard
form on $N$, then by the very definition of Hadamard form 
$w-\cG_n$ is given by a $C^n$-integral kernel, for all $n\in\bN$.
Therefore one obtains as in the proof of Thm.\ 5.1 in \cite{Rad1} that
${\rm WF}(w)\subset \cR\cap({\rm T}^*N\times{\rm T}^*N)$.

Denoting by $\omega_{2\; N\times N}$ the restriction of $\omega_2$ to 
$\Coin((\fV\bt\fV)_{N\times N})$, it follows that 
\begin{equation}
\label{equ32}
{\rm WF}(\omega_{2\; N\times N}) \subset \cR \cap({\rm T}^*N\times{\rm T}^*N)
\end{equation}
whenever $\omega_2$ is the two-point function of a Hadamard state on
$N$. (For the CCR case this is immediate as in
this case $\omega_{2\; N\times N}= w$ for some Hadamard form $w$ on
$N$. For the CAR case this follows since then 
$\omega_{2\; N\times N}=(D_{\dni}\otimes 1)w$ for some Hadamard form
$w$ on $N$, and application of differential operators cannot increase
the wavefront set.)

For any quasifree state $\omega$ fulfilling the CCR or CAR it
holds that $\o_2$ is a bisolution for the wave-operator (it would be
sufficient for the subsequent arguments that $\o_2$ be a bisolution
mod $\Cin$). 
This means that one can apply the PST, Prop.\ \ref{pro2}, in order to
show that \eqref{equ32} already implies
\begin{equation}
\label{equ33}
{\rm WF}(\omega_2)\subset\cR  
\end{equation}
owing to the fact that $N$ is a neighbourhood of a Cauchy-surface $\Sigma$:
Let $(q,\xi ;q',\xi')$ be an element of $\text{WF}(\omega_2)$. Then
the first part of Prop.\ \ref{pro2} shows that $\xi,\xi'$ are both 
lightlike. As any inextendible lightlike geodesic intersects 
$\Sigma$, we can --- provided that {\em both} $\xi$ and $\xi'$ are non-zero
--- use the second part of Prop.\ \ref{pro2} to conclude that 
$(p,\zeta ;p',\zeta')\in\text{WF}(\omega_2)$, where $(p,\zeta ;p',\zeta')$ is
the (unique) element of ${\rm B}(q,\xi)\times {\rm B}(q',\xi')$ with 
$p,p'\in \Sigma$.  
But then, because of \eqref{equ32}, $p=p', \zeta=-\zeta'$ with $\zeta'$
future pointing. Thus we conclude that $(q,\xi;q',\xi') \in \cR$.

 Now we will show that the PST in combination with the propagation of
Hadamard form entails that $(q,\xi,q',0)$ and $(q,0;q',\xi')$ are
absent from ${\rm WF}(\omega_2)$. 
We will give an indirect proof and thus assume that WF$(\o_2)$
contains an element of the form $(q,\xi;q',0)$. Then there will be a
Cauchy-surface $\Sigma'$ passing through $q'$; this Cauchy-surface
possesses a causal normal neighbourhood $N'$.
By the propagation of Hadamard form, $\o_2$, being a bisolution (mod
$\Cin$ would suffice) for the wave-operator, will be of Hadamard
form on $N'$. 
 Moreover, $\xi \ne 0$,
and so there is some point $(p,\zeta) \in {\rm B}(q,\xi)$, $\zeta \ne
0$, with $p \in \Sigma'$. Since $\o_2$ is of Hadamard form on $N'$, it
follows (see above) that WF$(\o_{2\,N'\times N'}) \subset \cR \cap
({\rm T}^*N' \times {\rm T}^*N')$, and thus $(p,\zeta;q',0)$ can 
only be contained in WF$(\o_2)$ if $p = q'$ and $\zeta = 0$. By the PST, this
contradicts the assumption that $(q,\xi;q',0) \in {\rm WF}(\o_2)$.

Thus elements of the form $(q,\xi;q',0)$ are absent from WF$(\o_2)$,
and by an analogous argument, also pairs of covectors of the form
$(q,0;q',\xi')$ aren't contained in WF$(\o_2)$. Thus we have
established the inclusion \eqref{equ33}.

Now we have to establish the reverse inclusion 
\begin{equation}
\label{equ34}
{\rm WF}(\omega_2)\supset \cR.   
\end{equation}
In order to prove this we use Prop.\ \ref{pro3} and Lemma \ref{lem5}
together with Proposition \ref{P-1}, showing that for any
Hadamard state $\omega$ on the CCR-algebra one has  
\begin{equation*}
{\rm WF}(\omega_2)\supset\left\{(q,\xi ;q,\xi')
\in {\rm T}^*_qN\times{\rm T}^*_qN \; :\; \xi \in \cN_-,\;
\xi'=-\xi\right\}\,.  
\end{equation*}
The same result can be derived for any Hadamard state $\omega$ 
on the CAR-algebra. 

According to the PST, this implies that 
\begin{equation*}
{\rm WF}(\omega_2)\supset {\rm B}(q,\xi)\times {\rm B}(q,\xi')
\end{equation*}
for all $\xi,\xi'\in {\rm T}_q^*N$ with $\xi\in\cN_-$, $\xi'=-\xi$.
Since this holds for all $q$ in the causal normal neighbourhood $N$ of
a Cauchy surface, relation \eqref{equ34} now follows. Thus we have
proved ${\rm WF}(\omega_2)=\cR$.
\\[6pt]
(b)  Now let $\omega$ be a state on the CCR or CAR algebra associated
to some (wave or Dirac) operator with the property that \eqref{6-WF}
holds. Let $N$ be a causal normal neighbourhood of any given
Cauchy-surface. According to Lemma \ref{lem4}, there is, in the
CCR-case, a Hadamard form $w$ on $N$ fulfilling \eqref{equ35}, and in
the CAR-case, there is a Hadamard form $w$ on $N$ obeying
\eqref{equ36}.
According to part (a) of the proof, it holds in either case, 
\begin{equation*}
{\rm WF}(w) =\cR \cap({\rm T}^*N\times{\rm T}^*N),  
\end{equation*}
so that one obtains
\begin{equation*}
{\rm WF}(w-\omega_{2\; N\times N}) 
\subset \cR \cap({\rm T}^*N\times{\rm T}^*N) \subset \cN_- \times \cN_+\,.  
\end{equation*}
Now we have in the CCR-case $w^{(-)}-\omega_{2\; N\times N}^{(-)}=0$.
Introducing the flip morphism
\begin{equation*}
\iota: M\times M \to  M\times M\,,\quad (p,q)\mapsto (q,p), 
\end{equation*}
and some bundle morphism $I$ covering $\iota$, 
as well as $u:=w-\omega_{2\; N\times N}$, this implies
\begin{equation*}
\text{WF}(u)=\text{WF}(u^{(+)})=\text{WF}(I^{\stern}u^{(+)})
={}^t\!D\iota\text{WF}(u^{(+)})= {}^t\!D\iota\text{WF}(u).
\end{equation*}
But because of the anti-symmetry of the set 
$\cN_-\times\cN_+$, its intersection with its image under
${}^t\!D\iota$, $\cN_+ \times \cN_-$, is
empty, so one finds   
\begin{equation*}
{\rm WF}(w-\omega_{2\; N\times N})=\emptyset. 
\end{equation*}
The same reasoning applies to the CAR-case, where 
$w^{(+)}-\omega_{2\; N\times N}^{(+)}=0$. Thus  
$\omega_2$ is shown to be of Hadamard form on $N$ in both cases.
This shows that $\o$ is a Hadamard state.  
\end{proof}

%
\begin{appendix}
\section{Appendix}
%
%
\subsection{The Hadamard coefficients}
%
\label{app.koeffs}
In this appendix we give the definition of the Hadamard coefficients
for the wave-operator $P$ on the vector-bundle $\fV$ according to
Chapter III in \cite{Gun}, adapted to our notation. 

Let $N$ be a causal normal neighbourhood of an arbitrary
Cauchy-surface, and let $\chi$ be an $N$-regularizing function with
support domain $\cU_*,\cU$. Then for $(x,y)\in\cU$, the (signed) square 
of the geodesic distance between $x$ and $y$, $s(x,y)$, is
well-defined and a smooth function of both arguments. For each
$(x,y)\in \cU$, denote by $\boldsymbol{s}_y(x)$ the vector
$\text{grad}_x\, s(x,y)$ in ${\rm T}_x N$, and define 
\begin{equation*}
M(x,y):= \frac{1}{2}\Box_x s(x,y)-m.   
\end{equation*}
Then by Prop. III.1.3 in \cite{Gun}, there is exactly one sequence
$\{U_{(k)}\}_{k\in\bN_0}$ of sections 
$U_{(k)}\in\Cin((\fV\bt \fV^{*})_{\cU})$ satisfying the differential
equations
\begin{equation}
\label{equ42}
(P\otimes 1)U_{(k-1)}(x,y)
+(\nabla_{\boldsymbol{s}_y(x)}^{(P)}\otimes1)U_{(k)}(x,y)
+(M(x,y)+2k)U_{(k)}(x,y)   
\end{equation}
with the initial conditions 
\begin{equation*}
U_{(-1)}(x,y)=0,\quad\quad U_{(0)}{}^a{}_b(x,x)=\delta^a{}_b,
\end{equation*}
where the latter condition is to be understood with respect to 
dual frame indices for $\fV$ and $\fV^*$, respectively, and the
differential operators in \eqref{equ42} act on the left tensor entry,
i.e. with respect to the variable $x$. (We caution the reader that at
this point our notation deviates from that in \cite{Gun}.) 

The members of the sequence $\{U_{(k)}\}_{k\in\bN_0}$ are called
{\it Hadamard coefficients}. 

With this definition, the sections $U$, $V^{(n)}$ and $T^{(n)}$ in 
$\Cin((\fV\bt \fV^{*})_{\cU})$ appearing in the main text (which are
also often referred to as Hadamard coefficients) are given by 
\begin{align*}
U(x,y)&:=\sum_{k=0}^{(m-4)/2}(4-m,k)^{-1}U_{(k)}(x,y)s(x,y)^k\\
V^{(n)}(x,y)&:=\left(2,\frac{m}{2}-1\right)\sum_{k=0}^n\frac{1}{2^kk!}
U_{\left((m-2)/2+k\right)}(x,y)s(x,y)^k\\
T^{(n)}(x,y)&:=\sum_{k=0}^{n + (m-3)/2}(4-m,k)^{-1}U_{(k)}(x,y)s(x,y)^k
\end{align*}
where the symbol $(\alpha,k)$ is defined by
\begin{equation*}
(\alpha,0)=1,\quad\quad (\alpha,k)=\alpha(\alpha+2)\ldots(\alpha+2k-2).
\end{equation*} 
If the wave-operator $P$ in question is symmetric w.r.t.\  the 
sesquilinear form \eqref{equ6-00}, the
corresponding Hadamard coefficients have an additional symmetry
property which we shall use below. To state this property, let
$\Theta=\vartheta\Gamma$ where $\vartheta$ and $\Gamma$ are the 
morphisms defined in the main text and define $\iota$ to be the flip
morphism 
\begin{align*}
\iota : \Cin(\fV\,\bt\,&\fV^*)\rightarrow \Cin(\fV^*\bt \fV),\\
&f(p)\otimes \n(q)\mapsto \n(q)\otimes f(p)\,,\quad
f\in\Cin(\fV),\;\n\in\Cin(\fV^*). 
\end{align*}
Note that the map $\Theta: \fV \to \fV^*$ induces a bilinear form
$\Theta({\rm v},{\rm v'}) = [\Theta {\rm v}]({\rm v'})$ on $\fV$, and
hence one can introduce its transpose $\Theta^T: \fV \to \fV^*$ by
$[\Theta^T {\rm v}]({\rm v'}) = \Theta({\rm v'},{\rm v})$. Using
frame-indices, we introduce the notation
\begin{equation}
\label{Utransp}
\big(\Theta^T U_{(k)} \Theta^{-1}\big)_a{}^b(p,q) = \Theta^T{}_{ac}(p)
U_{(k)}{}^c{}_d(p,q) (\Theta^{-1}){}^{db}(q)\,,
\end{equation}
thus defining $\Theta^T U_{(k)}\Theta^{-1} \in \Cin(\fV^* \bt \fV)$.
With that notation, we have 
\begin{Lemma}
\label{lem1}
If $P$ is symmetric, i.e.\ fulfills \eqref{equ6-a}, the section 
$ U_{(k)} -\iota\left(\Theta^T U_{(k)}\Theta^{-1}\right)$
vanishes faster than any power of $s(p,q)$ on the set 
$\{(p,q)\in\cU\,\vert\, s(p,q)=0\}$, for all
$k\in\bN$.  
\end{Lemma}
A proof of this fact can be obtained by combining the results of
Prop.\ 4.6 and 4.9 from \cite{Gun}, Chap. III. 
%
\subsection{Normal coordinates}
\label{app.nc}
We begin with some words on normal coordinates:
Let $\Omega$ be a convex normal neighbourhood, containing a point
$p$. Then we can cover $\Omega$ by normal coordinates
$\xi_p:\bR^m\rightarrow\Omega$ centered at $p$ as follows:
We identify $\bR^m$ and $\text{T}_p\Omega$, using a basis    
$v_{(k)}$ of $\text{T}_p\Omega$ which fulfills 
$g_p(v_{(i)},v_{(j)})=\eta_{ij}$, by
\begin{equation*}
w:\bR^m\mapsto\text{T}_p\Omega,\quad x=(x_0,\ldots,x_{m-1})
\mapsto w(x)=\sum_{i=0}^{m-1} x_iv_{(i)}
\end{equation*}
and let $\xi_p(x):=\exp_p(w(x))$ for all $x$.\\
A useful fact about normal coordinates is that 
\begin{equation}
\label{equ106}
-s(p,\xi_p(x))=\eta(x,x)\quad\text{ for }\ \ \xi_p(x)\in\Omega.
\end{equation}
Unfortunately, there is no such simple formula for the geodesic distance
$s(\xi_p(x),\xi_p(y))$ if both points are different from $p$.
There is, however, a useful approximation to it, which we will have
occasion to use below: Let $\lambda$ be in
$\bR$, $\lambda\geq 0$. Then we have
\begin{equation}
\label{equ107}
-s(\xi_p(\lambda x),\xi_p(\lambda y))=\lambda^2\eta(x-y,x-y)+
\lambda^4\phi(x,y),
\end{equation}
where $\phi(x,y)$ is a remainder which is smooth in $x,y$. 
For details see \cite{Syn}.

It will be useful to have a symbol for the pullback of functions $f$
in $\Coin(\Omega)$ via normal coordinates. We therefore define
\begin{equation*}
\acc{f}_p(x):=
\begin{cases}
f(\xi_p(x))\det(g_{\xi_p(x)})^{\frac{1}{2}}&\text{ for
  }\xi_p(x)\in\Omega\\
0&\text{ else}
\end{cases}.
\end{equation*}
\subsection{Some properties of $\chi G^{(\ez)}$}
\label{app.hadamard_form}
\label{app.dist}
We can now begin our investigation of the distributions showing up in our
definition of a Hadamard form by considering special kinds of
distributions on Minkowski space: 

Let $t$ be a time function on $m$ dimensional Minkowski space with
$t(0)=0$ and define 
\begin{equation}
\label{equ100}
\tilde {G}^{(\ez)} (f)=\lim_{\varepsilon\rightarrow 0+}
\int f(x)G^{(\ez)}_{\varepsilon}(0,x)\;
\text{d}^m x \,,\quad f\in\Coin(\bR^m)\,,
\end{equation}
where the functions $G^{(1)}_{\varepsilon}, G^{(2)}_{\varepsilon}$
are given by equations  
\eqref{equ101.a}, \eqref{equ102}, taken in the case of the Minkowski metric
(i.e.\ $-s(x,y)=\eta_{\m \n}(x-y)^{\m}(x-y)^{\n} = \eta(x-y,x-y)$)
 and with $\chi\equiv 1$. 
Although the time function $t$ enters the
above definition, the distributions $\tilde{G}^{(1)}$ and $\tilde{G}^{(2)}$
do actually not depend on $t$. More precisely, we have
\begin{Lemma}
\label{lem2} 
\begin{equation}
\label{equ104}
\begin{split}
\tilde{G}^{(1)}(f)&=\begin{cases}
\beta(m+1,m)\int
\big[\theta(-\eta(x,x))\sqrt{-\eta(x,x)}\\
\quad\quad\quad\quad\quad -i\sign(x_0)\theta(\eta(x,x))\sqrt{\eta(x,x)}\big]
\Box^{\frac{m-1}{2}}f(x)\;\text{d}^{m}x
&\text{ for $m$ odd}\\
\beta(m+2,m)\int\big[\frac{1}{\pi}\eta(x,x) \ln \betr{\eta(x,x)}\\
\quad\quad\quad\quad\quad -i\sign(x_0)\theta(\eta(x,x))\eta(x,x)\big]
\Box^{\frac{m}{2}}f(x)\;\text{d}^{m}x&\text{ for $m$ even} 
\end{cases}\\
\tilde{G}^{(2)}(f)&=
\beta(m+2,m)\int\left[\frac{1}{\pi}\eta(x,x) \ln \betr{\eta(x,x)}
  -i\sign(x_0)\theta(\eta(x,x))\eta(x,x)\right]\times \\
& {} \hspace*{4.5cm}\times (\Box+1+2/m)
f(x)\;\text{d}^{m}x
\end{split}
\end{equation}
where $\theta(s) =1$ for $s \ge 0$ and $\theta(s) = 0$ for $s < 0$, and 
\begin{equation*}
\beta(\alpha,m):= 2^{1-\alpha}\pi^{(2-m)/2}
\left[\Gamma\left(\frac{\alpha-m}{2}+1\right)
\Gamma\left(\frac{\alpha}{2}\right)\right]^{-1}. 
\end{equation*}
\end{Lemma}
\begin{proof}
As first step, we will prove independence of $t$   
by generalizing an argument given in \cite{KW} to arbitrary
dimensions:
 
For $\tilde{G}^{(2)}$ note that $G^{(2)}_{\epsilon}$ converges for 
$\varepsilon\rightarrow 0$ to a locally integrable function which
does not depend on $t$ anymore. As we can use Lemma B2 of \cite{KW} to
conclude that we may interchange integration and limit in
\eqref{equ100}, we see that $\tilde{G}^{(2)}$ is indeed a well defined
distribution and independent of $t$.

The limit of $G^{(1)}_{\varepsilon}$ for $\varepsilon\rightarrow 0$
is not locally integrable, so before we can apply Lemma B2 of
\cite{KW} to argue as above, we will have to rewrite \eqref{equ100},
using integration by parts. 
To this end, introduce 
coordinates $\tau,\sigma,\vartheta$
on $\bR^{m}$, where 
\begin{equation*}
\tau(x_0,\ldots,x_{m-1})=x_0,\quad \quad \sigma
(x_0,\ldots,x_{m-1})=\sum_{i=1}^{m-1}x_i^2
\end{equation*}
and $\vartheta$ stands for some coordinatization of
$S^{m-2}$. In these coordinates
\begin{equation*}
\tilde{G}^{(1)} (f)=\lim_{\varepsilon\rightarrow 0+}
\iiint \text{d}\tau\,\text{d}\sigma \,\text{d}\vartheta\;
f(\tau,\sigma ,\vartheta)
\sigma ^{\frac{m-3}{2}}\left(\sigma -\tau^2 -i2\varepsilon 
t(\tau,\sigma ,\vartheta) +\varepsilon^2
\right)^{-\frac{m}{2}+1}
\end{equation*}
In case $m$ is even, we carry out a $m/2-1$-fold partial
integration with respect to $\sigma$ and arrive at 
\begin{equation*}
\begin{split}
\tilde{G}^{(1)}(f)=c\lim_{\varepsilon\rightarrow 0+}
\iiint \text{d}\tau\,\text{d}\sigma \,\text{d}\vartheta\;
\ln&\left(\sigma-\tau^2 -i2\varepsilon t+\varepsilon^2\right)\\
&\Big(\,\underset{\frac{m}{2}-1\text{ times}}
{\underbrace{\partial_\sigma \big(\frac{1}{1-i2\varepsilon \partial_\sigma 
t}\partial_\sigma \big( \frac{1}{1-i2\varepsilon \partial_\sigma  t}\ldots}}
\;\sigma^{\frac{m-3}{2}}f\big)\big)\Big),
\end{split} 
\end{equation*}
where $c$ is some constant.
Now, the limit of the integrand is locally integrable and turns out to
be independent of $t$, too. Hence Lemma B2 of \cite{KW} can be applied
to show the desired result.

In case $m$ is odd, we carry out a $(m-3)/2$-fold partial integration with
respect to $\sigma$ and arrive at
\begin{equation*}
\begin{split}
\tilde{G}^{(1)}(f)=c'\lim_{\varepsilon\rightarrow 0+}
\iiint \text{d}\tau\,\text{d}\sigma \,\text{d}\vartheta\;&
\left(\sigma -\tau^2 -i2\varepsilon t+\varepsilon^2\right)^{-1}\\
\left(\sigma -\tau^2 -i2\varepsilon t+\varepsilon^2\right)^{\frac{1}{2}}
&\Big(\,\underset{\frac{m-3}{2}\text{ times}}
{\underbrace{\partial_\sigma \big(\frac{1}{1-i2\varepsilon \partial_\sigma 
t}\partial_\sigma  \big(\frac{1}{1-i2\varepsilon \partial_\sigma  t}\ldots}}
\;\sigma^{\frac{m-3}{2}}f\big)\big)\Big),
\end{split}
\end{equation*}
where again $c'$ is a suitable constant.
Since the limit of the integrand is not locally integrable in $\tau$, we are
not ready to apply Lemma B2 of \cite{KW} yet. But we have already
written the integrand in a suggestive form as to the next partial
integration. This turns $(\sigma-\tau^2 -i2\varepsilon
t+\varepsilon^2)^{-1}$  into a logarithm and
the $(\sigma -\tau^2  -i2\varepsilon t_p+\varepsilon^2)^{1/2}$ to
$(\ldots)^{-1/2}$, at worst, thus rendering the integrand locally
integrable in the
limit $\varepsilon\rightarrow 0$. We also get a boundary term at the
integration boundary $\sigma=0$, which is integrable in the limit
$\varepsilon\rightarrow 0$ as well. Inspection of these
limits shows that they are indeed independent of $t$,
whence $G^{(1)}$ is well defined and independent of $t$ also in this
case. 

As a consequence of the independence of $t$, we can write the
distributions in the form given in the statement of the Lemma: 
Using the trivial time-function 
$t_0(x):=x_0$ and the abbreviation
$g_\varepsilon(x):=-\eta(x,x)-i2\varepsilon x_0+\varepsilon^2$, we get 
\begin{align*}
\tilde{G}^{(1)}(f)&=\lim_{\varepsilon\rightarrow 0}
\begin{cases}
\beta(m+1,m)\int\sqrt{g_\varepsilon}(x)\Box^{\frac{m-1}{2}}f(x)
\;\text{d}^{m}x&\text{ for $m$ odd}\\  
-\frac{1}{\pi}\beta(m+2,m)\int g_\varepsilon \ln(g_\varepsilon)(x)
\Box^{\frac{m}{2}}f(x)\;\text{d}^{m}x&\text{ for $m$ even}
\end{cases}\\
\tilde{G}^{(2)}(f)&=-\frac{1}{\pi}\beta(m+2,m)\lim_{\varepsilon\rightarrow
  0}\int g_\varepsilon \ln(g_\varepsilon)(x)(\Box+1+2/m)
f(x)\;\text{d}^{m}x 
\end{align*}
By exchanging the integration and the limit, we finally get the
desired result. 
\end{proof}
We also have to introduce the so called \textit{Riesz distributions}, as
defined in \cite{Gun}, Chap. II. To this end, let $\alpha>m$ 
and define the distributions
\begin{equation*}
\tilde{R}(\alpha)[f]=-\beta(\alpha,m)\int
\sign(x_0)\theta(\eta(x,x))(\eta(x,x))^{\frac{\alpha-m}{2}}f(x)\;\text{d}^m
x,\quad f\in\Coin(\bR^n). 
\end{equation*} 
Note that $\beta$ is chosen such that 
$\tilde{R}(\alpha)[\phi]=\tilde{R}(\alpha+2)[\Box \phi]$
for all $\alpha>m$. We \textit{define} the distributions
$\tilde{R}(\alpha)$ 
for \textit{all} $\alpha\in\bR$ by means of this relation.

As before, let $\Omega$ denote a causal normal neighbourhood and
$\xi_p$ normal coordinates on $\Omega$. We can then define
distributions $R^\Omega(\alpha)$ on
  $\Coin(\Omega\times\Omega)$ by 
\begin{equation*}
R^\Omega(\alpha)[f\otimes f']:=
\int f(p) \tilde{R}(\alpha)[\acc{f'}_p]\;\text{d}\mu(p),\quad\quad
f,f'\in\Coin(\Omega)   
\end{equation*} 
and continuous extension to $\Coin(\Omega\times\Omega)$. These so
called \textit{Riesz distributions} bear a certain relation to the
distributions we are really interested in:
\hide{
We also wish to introduce the antisymmetric part of a distribution:
Let $G$ be a distribution on $\Coin(\Omega\times\Omega)$. We define
the antisymmetric part of $G$ by
\begin{equation*}
G^{(-)}(f\otimes f') := \frac{1}{2}\left(G(f\otimes f')-G(f'\otimes f)\right)
\quad\quad f,f'\in\Coin(\Omega)
\end{equation*}
and continuous extension.} 
Let $\Omega$ be a normal neighbourhood, $t$ be a time function on
$\Omega$ and denote by $G^{(1,\Omega)}$, $G^{(2,\Omega)}$
the distributions on $\Coin(\Omega\times\Omega)$ obtained by setting 
$\chi\equiv 1$ in the definition of the distributions 
$\chi G^{(1)}$, $\chi G^{(2)}$, respectively. We caution the reader,
that, as a consequence, $G^{(1,\Omega)}$ and $G^{(2,\Omega)}$ are
not well defined on $\Coin(M\times M)$ in contrast to $\chi G^{(1)}$, 
$\chi G^{(2)}$. 

Using normal coordinates on $\Omega$ (especially their property
\eqref{equ106}) and interchanging limit and
integration, we see that
\begin{equation*} 
G^{(\ez,\Omega)}(f\otimes f') 
=\int f(p)\tilde{G}^{(\ez)}(\acc{f'}_p)\;\text{d}p,\quad\quad 
f,f'\in\Coin(\Omega).
\end{equation*}
The interchange of limit and integration is valid because the 
limit $\varepsilon\rightarrow 0$ in $\tilde{G}^{(1/2)}(\acc{f'}_p)$
is uniform in $p$ on compact sets. Thus we see that $G^{(1,\Omega)}$,
$G^{(2,\Omega)}$ are actually 
independent of $t$. Moreover, by comparison with the definition of
$R(\alpha)$, using the formulae \eqref{equ104}, we find that 
\begin{equation}
\label{equ106.a}
G^{(1,\Omega)(-)}=iR^\Omega(2),\quad\text{ and }\quad
G^{(2,\Omega)(-)}=iR^\Omega(m). 
\end{equation}
\subsection{Proof of Lemma 5.4}
\label{app.cons}
(i) {\it Existence of Hadamard forms}\\[6pt]
Following Sec.\ 4.3 in \cite{Fri}, the argument that Hadamard forms
for the wave-operator $P$ exist at all on a causal normal
neighbourhood $N$ runs as follows:

Assume that $\phi \in \Coin(\bR)$ has the following properties: $0 \le
\phi(s) \le 1$, $\phi(s) = 1$ for $|s| \le 1/2$, and $\phi(s) = 0$ for
$|s| \ge 1$. Then one can show (cf.\ \cite[Lemma 4.3.2]{Fri},
\cite[Prop.\ III.2.6.3]{Gun}) that
there exists a strictly increasing and diverging sequence
$(\k_j)_{j\in \bN}$ of natural numbers so that the modified Hadamard
coefficient sections $\tilde{V}^{(n)}$ and $\tilde{T}^{(n)}$, which are
defined like $V^{(n)}$ and $T^{(n)}$, but with the terms $s(x,y)^k$ replaced
by $s(x,y)^k\cdot \phi(\k_k s(x,y))$, converge for $n \to \infty$
uniformly on compact 
subsets of $\cU$ to smooth sections $\tilde{V}$ and $\tilde{T}$,
respectively. Moreover, it holds that for all $n$,
$$ s(x,y)^{-n}(\tilde{V}(x,y) - V^{(n)}(x,y)) \quad {\rm and} \quad
s(x,y)^{-n}(\tilde{T}(x,y) - T^{(n)}(x,y)) $$
converge to $0$ as $s(x,y) \to 0$. Thus it is not difficult to check
that (for $m$ even)
$f,f' \mapsto\chi G^{(2)}((\vartheta f)(V^{(n)} - \tilde{V})f')$ is
given by a $C^n$-kernel, and likewise (for $m$ odd)
$f,f' \mapsto \chi G^{(1)}((\vartheta f)(T^{(n)} - \tilde{T})f')$ is
given by a $C^n$-kernel. This guarantees the existence of Hadamard
forms on $N$.
\\[6pt]
(ii) In a next step, one must show that the $H^{(n)} \in C^n((\fV \bt
\fV^*)_{N \times N})$ can be chosen
such that (5.4), respectively (5.5), are fulfilled. Let us treat the
case of relation (5.4) first.

We note that in case $p,q \in N$ lie acausal to each other, there is a
neighbourhood $\cV$ in $N \times N$ such that $E$ vanishes on $\cV$,
and any Hadamard form on $\cV$ is $\Cin$. One can thus correct the
Hadamard form $w$ on $\cV$ by a $\Cin$ integral kernel so as to obtain
a new Hadamard form vanishing
on $\cV$. Hence, we need only consider the situation in a
neighbourhood of causally related points $(p,q) \in N \times N$ (and
eventually use a partition of unity argument). Given a pair of
causally related points $p,q \in N$, there is, by definition of causal
normal neighbourhood, a convex normal neighbourhood $\O$ containing
both $p$ and $q$.
  
The importance of the distributions $R^\Omega(\alpha)$ defined above 
lies in the fact that
they show up in the explicit formula for the fundamental solution $E^\Omega$
of a wave-operator $P$ on $\Omega$, given in \cite{Gun}:
\begin{equation}
\label{equ105}
\scprr{f}{E^{\Omega}f'}=\begin{cases}
R^\Omega(2)\left[\vartheta(f)T^{(n)}f'\right]+\scprr{f}{M^{(n)}f'}&\text{
  for $m$ odd}\\
R^\Omega(2)\left[\vartheta(f)Uf'\right]
+R^{\Omega}(m)U\left[\vartheta(f)V^{(n)}f'\right] 
+\scprr{f}{M^{(n)}f'}&\text{ for $m$ even}
\end{cases}
\end{equation}
where $U_{(k)}$, $T^{(n)}$, $V^{(n)}$ and $U$ are the sections defined
in Appendix \ref{app.koeffs} and 
$M^{(n)}\in C^{n}((\fV \bt \fV^*)_{\O \times \O})$ are suitably chosen.
We have also used the shorthands
\begin{eqnarray*}
\vartheta(f)T^{(n)}f'(p,q)&:=&\vartheta(f)_a(p)T^a{}_b(p,q){f'}^b(q)\,,\\
\quad (f,M^{(n)}f')& := &\int (\vartheta f)_a(p) M^a{}_{b}(p,q)f'{}^b(q)
\, d\m (q) d\m (p)\,, \quad {\rm etc.}
\end{eqnarray*}

Now let $w$ be a Hadamard form, and let $f,f' \in \Coin(\fV_{\O})$. Then
\begin{equation*}
\begin{split}
w^{(-)}&(\Gamma f\otimes f')=\\
&\frac{1}{2}\begin{cases}
  G^{(1,\Omega)}(\vartheta(f)\chi T^{(n)}f')
+\scprr{f}{H^{(n)}f'}\\
\quad-G^{(1,\Omega)}(\vartheta(\Gamma f')\chi T^{(n)}\Gamma f)
-\scprr{\Gamma f'}{H^{(n)}\Gamma f}&\text{ for $m$ odd}\\
G^{(1,\Omega)}(\vartheta(f)\chi U f')+G^{(2,\Omega)}
(\vartheta(f)\chi V^{(n)}f')+\scprr{f}{H^{(n)}f'}\\
\quad-G^{(1,\Omega)}(\vartheta(\Gamma f')\chi U \Gamma (f))
-G^{(2,\Omega)}(\vartheta(\Gamma f')\chi V^{(n)} \Gamma f)
-\scprr{\Gamma f'}{H^{(n)}\Gamma f}&\text{ for $m$ even}
\end{cases}
\end{split}
\end{equation*}
Using the fact that $\Gamma$ is a conjugation in the CCR case, one
can compute that (cf.\ \eqref{Utransp} for notation)
\begin{equation*}
G^{(1,\Omega)}(\vartheta\Gamma(f')\chi T^{(n)}\Gamma (f))=
G^{(1,\Omega)}(\iota(\vartheta (f)\iota(\Theta^T \chi T^{(n)}\Theta^{-1})f'))
\end{equation*}
Now we can use the fact that $\chi$ is identically 1 on 
$\{(p,q)\in\Omega\times\Omega\,\vert\, s(p,q)=0\}$ together with
Lemma \ref{lem1} to the effect that   
\begin{equation*}
G^{(1,\Omega)}(\vartheta\Gamma(f')\chi T^{(n)}\Gamma (f'))
=G^{(1,\Omega)}(\iota(\vartheta(f)\chi T^{(n)}f')) \text{ mod } C^{\infty}\,. 
\end{equation*}
Treating the other terms with a minus sign in a similar way, we arrive
at
\begin{equation*}
\begin{split}
w^{(-)}&(\Gamma f\otimes f')=\\
&\begin{cases}
G^{(1,\Omega)(-)}(\vartheta(f)\chi T^{(n)}f')+\scprr{f}{H^{(n)}f'}\\ 
\quad-\scprr{f}{\iota(\Theta^T H^{(n)}\Theta^{-1})f'}
+\scprr{f}{K^{(n)}f'}&\text{ for $m$ odd}\\
G^{(1,\Omega)(-)}(\vartheta(f)\chi U f')+G^{(2,\Omega)(-)}
(\vartheta(f)\chi V^{(n)}f')+\scprr{f}{H^{(n)}f'}\\
\quad -\scprr{f}{\iota(\Theta^T H^{(n)}\Theta^{-1})f'}
+\scprr{f}{K^{(n)}f'}&\text{ for $m$ even}
\end{cases}
\end{split}
\end{equation*}  
where $K^{(n)}\in C^{\infty}((\fV \bt \fV^*)_{N \times N})$ 
reflects the potential asymmetry of
$\chi,U_{(k)}$ away from 
$\{(p,q)\in\Omega\times\Omega\,\vert\, s(p,q)=0\}$. Upon
using the identification  \eqref{equ106.a} between $R^\Omega(\alpha)$ and 
$G^{(...,\Omega)(-)}$ and comparing with the expression \eqref{equ105}, 
one can now see that it is indeed possible to choose the $H^{(n)}$ in 
such a way that $w^{(-)}(\Gamma f\otimes f')=i(f, E^\Omega
f')$. Because of uniqueness of the fundamental solution, we have 
$E^\Omega=E\vert_\Omega$, which concludes the proof of the lemma in the
CCR case. 

For the CAR case, note that $\Gamma$ acts as a
skew-conjugation. Thus, the computation can be carried through as
above, and one can choose the sections ${H'}^{(n)}$ of $w'$ such that 
\begin{equation*} 
{w'}^{(+)}(\Gamma D_{\dni}f\otimes f')=i\scprr{D_{\dni}f}{Ef'}\,,\quad
f,f'\in\Coin(\fV_N)\,,
\end{equation*}
which in turn gives the desired result. 
\\[6pt]
(iii)
Finally, we have to show that part (c) of Lemma 5.4 holds. We follow
the argument given in the ``Note added in proof'' in \cite{Rad1}. To this end
we note that by part (a), $w^{(-)}$, the antisymmetric part of a
Hadamard form $w$, is always a bisolution mod $\Cin$ for the
wave-operator $P$. On the other hand, according to Thm.\ 5.1, part (i),
in \cite{Rad1} (cf.\ the proof of our Thm.\ 5.8), it holds that WF$(w)
\subset \cR$, where the set $\cR$ has been defined in (5.9); it is
significant that $\cR \subset \cN_- \times \cN_+$. Now since $w^{(-)}$
is a bisolution mod $\Cin$ for the wave-operator $P$, it holds that
WF$(w^{(P)(-)}) = \emptyset$, where $w^{(P)}$ has been defined in
Sec.\ 3.2. Thus WF$(w^{(P)}) = {\rm WF}(w^{(P)(+)}) \subset \cN_- \times
\cN_+$. But since $w^{(P)(+)}$ is symmetric, its wavefront set must be
invariant under ${}^t\!D\iota^{-1}$ where $\iota : (p,q) \mapsto (q,p)$ is
the `flip' morphism on $N \times N$. That is, one concludes exactly as
in part (b) of the proof of Thm.\ 5.8 that WF$(w^{(P)(+)})$ must be
contained in $(\cN_+ \times \cN_-) \cap (\cN_- \times \cN_+) =
\emptyset$. Similarly one concludes that WF$(w_{(P)}{}^{(+)})$ is
empty, and thus $w$ is a bisolution up to $\Cin$ for the wave-operator $P$.  
\subsection{Scaling limits}
\label{app.scale}
In this section, we will prove Prop.\ \ref{pro3} of the main text,
determining the scaling limit of a Hadamard distribution. Let, for the
rest of the section, $p$ be some point of $M$ and $\Omega$ a convex
normal neighbourhood of $p$, small enough such that $\fV_\Omega$
trivializes. 
Let the morphisms $\delta_\lambda$, 
$D_\lambda^{(\alpha)}$ be defined as in Section 
\ref{sec55}. Additionally define the action of dilations on test
functions $f\in\Coin(\Omega)$ by 
$d_\lambda^{(\alpha)}f=\lambda^{-\alpha}f\circ\delta^{-1}_\lambda$.
We will also use the shorthand $G^{(1)}_{\eta}$ for the
distribution $G^{(1,\Omega')}$, evaluated on Minkowski space (i.e. $g=\eta$,
$\Omega'=\bR^m$).  

As the main step of the proof, we will compute the scaling limit of
the distributions $G^{(1,\Omega)}$, $G^{(2,\Omega)}$:
\begin{Lemma}
Let $\alpha=m/2+1$ and $F \in \Cin(\O \times \O)$. Then for all
$f,f'\in \Coin(\O)$ there holds 
\begin{eqnarray*}
\lim_{\lambda\rightarrow 0}G^{(1,\Omega)}(F\cdot(d_\lambda^{(\alpha)}f \otimes
d_\lambda^{(\alpha)}f'))&=&F(p,p)\cdot G_{\eta}^{(1)}(f\lcrc\xi_{p}\otimes
f'\lcrc\xi_{p})\,,\\
\lim_{\lambda\rightarrow
  0}G^{(2,\Omega)}(F\cdot(d_\lambda^{(\alpha)}f \otimes 
d_\lambda^{(\alpha)}f'))&=&0. 
\end{eqnarray*}
\end{Lemma}
\begin{proof}
Since for each $f \in \Coin(\O)$ the support of $d_{\l}^{(\a)}f$ is for
$\l \to 0$ shrinking to $p$, it suffices to prove the statement for
the case that $F\in\Coin(\O \times \O)$. We will demonstrate 
the statement only for simple tensors $F = u \otimes u'$
with $u,u' \in \Coin(\O)$ since this results in slightly simpler
notation, but it will be obvious from the argument that general $F \in
\Coin(\O \times \O)$ can be dealt with in exactly the same manner.
 
We begin by considering $G^{(1,\Omega)}$ in case $m$ is odd: 
By using the results of App.\ \ref{app.dist}, we see that 
\begin{equation*}
G^{(1,\Omega)}(ud_\lambda^{(\alpha)}f \otimes u'd_\lambda^{(\alpha)}f')=
c\iint(ud_\lambda^{(\alpha)}f)(q)h(v)\Box^{\frac{m-1}{2}}
(u'd_\lambda^{(\alpha)}f')\acc{}_q(v)\;\text{d}^mv\,\text{d}\mu(q)
\end{equation*}
where $c$ is some constant depending on $m$ and 
\begin{equation}
\label{equ108}
h(v)=\theta(-\eta(v,v))\sqrt{-\eta(v,v)}
-i\sign(v_0)\theta(\eta(v,v))\sqrt{\eta(v,v)}.  
\end{equation}
Now we change the integration variables from $(q,\xi^{-1}_q(q'))$ to
$(\xi^{-1}_{p}(q),\xi^{-1}_{p}(q'))$, which we denote by
$(x,y)$. We get
\begin{align*}
&G^{(1,\Omega)}(ud_\lambda^{(\alpha)}f \otimes u'd_\lambda^{(\alpha)}f')\\
&=c\iint\lambda^{-2\alpha}u(\xi_p(x))f(\xi_{p}(\frac{x}{\lambda}))
h(v(x,y))\cdot \\
& {}\quad \quad \quad \quad \cdot \left(\Box^{\frac{m-1}{2}}_y[u'(\xi_p(y))
f'(\xi_{p}(\frac{y}{\lambda}))] 
+\bla\right)
\gamma_{\xi_{p}(x)}^{\frac{1}{2}}\gamma_{\xi_{p}(y)}^{\frac{1}{2}}
\;\text{d}^{m}y\,\text{d}^mx\\ 
&=c\iint \lambda^{m-2}u(\xi_p(\l x))f(\xi_{p}(x))h(v(\lambda x,\lambda y))\cdot
\\
& {} \quad \quad \quad \quad 
\cdot\left(\lambda^{1-m}\Box^{\frac{m-1}{2}}_y[u'(\xi_p(\l y))
f'(\xi_{p}(y))]+O(\lambda^{2-m})
\right)\gamma_{\xi_{p}(\lambda x)}^{\frac{1}{2}}\gamma_{\xi_{p}(\lambda
  y)}^{\frac{1}{2}} \;\text{d}^{m}y\,\text{d}^mx
\end{align*}
where $\gamma_p=\betr{\det(g_p)}$.
To compute the limit $\lambda\rightarrow 0$, we have to investigate
the behaviour of $\lambda^{-1}h(v(\lambda x,\lambda y))$. We use equation 
\eqref{equ107} and the fact that 
\begin{equation*}
-v_0(x,y)=x_0-y_0 +O(s(q,q'))+O(s(p,q'))+O(s(p,q))  
\end{equation*}
where we have set $q=\xi_{p}(x)$,
$q'=\xi_{p}(y)$ (see \cite[Chp.\ II,\S 9]{Syn} for details), to conclude 
\begin{equation*}
\lim_{\lambda\rightarrow 0}\lambda^{-1}h(v(\lambda x,\lambda y))=h(x-y)
\end{equation*}
with $h$ given by \eqref{equ108}. Using this, and
$\lim_{\lambda\rightarrow
  0}\det(g_{\xi_{p}(\lambda x)})=\det(g_{p})=1$ we get  
the desired result. 

The case $m$ even is treated exactly the same way, with the exception
that there is a term $(x-y)^2\ln\lambda^2$ in 
$\lambda^{-1}h(v(\lambda x,\lambda y))$ which seems to
blow up for $\lambda\rightarrow 0$. Using partial integration, it is
easy to see, though, that the term in $G^{(1,\Omega)}$ resulting from this
term in $h$ vanishes for any $\lambda$. Therefore, the rest of the argument    
goes through unchanged. 

The argument for $G^{(2,\Omega)}$ runs along the same lines.  
\end{proof}
Proposition \ref{pro3} is now a corollary of the above Lemma:
Let $U$ be in $C^{\infty}(\fV\bt \fV^*)$. In the components of the
frame $(e_i)$ used to define $D_{\l}$ in \eqref{dil}, we write
$$ [(\vartheta \lcrc\G D^{(\a)}_{\l})U D^{(\a)}_{\l}f'](q,q') =
 \Theta_{ac}(q)(D^{(\a)}_{\l}f)^c(q)U^a{}_b(q,q')(D^{\a}_{\l}f')^b(q')
 $$
and thus we obtain by the lemma,
$$
\lim_{\lambda \rightarrow 0}
G^{(1,\Omega)}\left(((\vartheta\lcrc\G D_\lambda^{(\alpha)}f)
UD_\lambda^{(\alpha)}f')\right)
= G^{(1)}_{\eta}\left((\boldsymbol{\Theta}R^{\stern}f)\boldsymbol{U}
R^{\stern}f'\right)\,,$$
where $\boldsymbol{U}$ denotes
the image of $U\vert_{(p,p)}$ under $R\otimes R$. 
The content of Prop.\ \ref{pro3}
concerning the CCR-case is now a consequence of simple properties of 
the Hadamard coefficients, such as $U_{(0)}(p,p)=1$.
In the CAR case, the appearance of an 
additional factor $\lambda^{-1}$ due to the differential operator 
$D_{\dni}$ has to be  compensated by a different choice of 
$\alpha$, as done in the proposition.     
\end{appendix}
%

\end{document}